\documentclass[conference]{IEEEtran}
\IEEEoverridecommandlockouts
\usepackage{cite}
\usepackage{amsmath,amssymb,amsfonts,hhline}
\usepackage{algorithmic}
\usepackage{graphicx}
\usepackage{textcomp}
\usepackage{xcolor}
\usepackage{booktabs} % For formal tables
\usepackage{xcolor}
\usepackage{soul}
\usepackage[utf8]{inputenc}
\usepackage{caption}
\usepackage{makecell}
\usepackage{amsmath,bm,amsfonts,graphicx,enumerate,multirow,array,tikz}
\usepackage{mathtools}
\usepackage{footmisc}
\usepackage{subcaption,graphicx}
\usepackage{changepage,enumitem,hyperref}
\usepackage{romannum}
\newcommand{\norm}[1]{\left\lVert#1\right\rVert}
\newcommand{\propose}{\textsf{{TransCF}}}
\newcommand{\proposealt}{\propose$^{\mathrm{alt}}$}
\newcommand{\proposedot}{\propose$^{\mathrm{dot}}$}
\newcommand{\proposeemb}{\propose$^{\mathrm{emb}}$}
\newcommand{\etal}{\textit{et al}. }

\def\BibTeX{{\rm B\kern-.05em{\sc i\kern-.025em b}\kern-.08em
    T\kern-.1667em\lower.7ex\hbox{E}\kern-.125emX}}
\begin{document}
\newcolumntype{C}{>{\centering\arraybackslash}p{4em}}
\newcolumntype{I}{>{\centering\arraybackslash}p{4.7em}}
\newcolumntype{P}{>{\centering\arraybackslash}p{2.9em}}
\newcolumntype{Y}{>{\centering\arraybackslash}p{3em}}
\newcolumntype{B}{>{\centering\arraybackslash}p{3.5em}}
\newcolumntype{D}{>{\centering\arraybackslash}p{3.8em}}
\newcolumntype{S}{>{\centering\arraybackslash}p{1.4em}}
\newcolumntype{L}{>{\centering\arraybackslash}p{1.8em}}
\newcolumntype{Q}{>{\centering\arraybackslash}p{4em}}
\newcolumntype{Z}{>{\centering\arraybackslash}p{2.3em}}
\newcolumntype{N}{>{\centering\arraybackslash}p{2.0em}}
\newcolumntype{A}{>{\centering\arraybackslash}p{2.2em}}
\newcolumntype{H}{>{\centering\arraybackslash}p{2.5em}}
\newcolumntype{J}{>{\centering\arraybackslash}p{2.7em}}
\title{Collaborative Translational Metric Learning}

%email address}}
%\thanks{$^*$Corresponding author}
\author{\IEEEauthorblockN{Chanyoung Park$^1$, Donghyun Kim$^2$, Xing Xie$^3$, Hwanjo Yu$^{1*}$\thanks{*Corresponding Author}}
\IEEEauthorblockA{\textit{Dept. of Computer Science and Engineering, POSTECH, South Korea$^{1}$} \\
\textit{Oath, USA$^{2}$}\\
\textit{Microsoft Research Asia, China$^{3}$}\\
\{pcy1302, hwanjoyu\}@postech.ac.kr, cartopy@gmail.com, xingx@microsoft.com}
%\and
%\and
%\and
%\and
%\and
}

\maketitle

\begin{abstract}
Recently, matrix factorization--based recommendation methods have been criticized for the problem raised by the triangle inequality violation.
Although several metric learning--based approaches have been proposed to overcome this issue, existing approaches typically project each user to a single point in the metric space, and thus do not suffice for properly modeling the \textit{intensity} and the \textit{heterogeneity} of user--item relationships in implicit feedback. In this paper, we propose~\propose~to discover such latent user--item relationships embodied in implicit user--item interactions. Inspired by the translation mechanism popularized by knowledge graph embedding, we construct user--item specific translation vectors by employing the neighborhood information of users and items, and translate each user toward items according to the user's relationships with the items. Our proposed method outperforms several state-of-the-art methods for top-N recommendation on seven real-world data by up to 17\% in terms of hit ratio. We also conduct extensive qualitative evaluations on the translation vectors learned by our proposed method to ascertain the benefit of adopting the translation mechanism for implicit feedback--based recommendations.
\end{abstract}

\begin{IEEEkeywords}
Recommender system, Metric learning, Collaborative filtering
\end{IEEEkeywords}

%\vspace{-1.5ex}
\section{Introduction}
%%\iffalse
%The recent exploding growth of information on the Web inundates users with choices, and recommender systems are playing a crucial role not only in helping users' decision-making process, but also in increasing the revenues of e-commerce companies.
%According to the recent technical report from Amazon.com~\cite{amazon}, estimated 30 percent of their page views were from recommendations, and likewise as for Netflix, more than 80 percent of movies watched on Netflix came through recommendations, and the value of Netflix recommendations is estimated at more than US\$1 billion per year~\cite{netflix}.
%As a consequence, a plethora of research has been devoted to building successful recommender systems.
%improving recommendation quality.
%\fi

The recent explosive growth of information on the Internet inundates users with choices, and recommender systems play a crucial role not only in helping users in their decision making, but also in increasing the revenues of e-commerce companies.
Among various recommendation techniques, matrix factorization (MF)--based collaborative filtering (CF)~\cite{koren2008factorization} has been shown to be the most successful; it assumes that users who have had similar interests in the past will tend to share similar interests in the future~\cite{bobadilla2013recommender}. More specifically, MF learns low-rank vector representations of users and items from their previous interaction history and models the similarity of user--item pairs using inner products. However, a critical yet not widely recognized flaw of employing the inner product as a similarity metric is that it violates the triangle inequality~\cite{ram2012maximum}. As a concrete example, if user $u_1$ liked items $v_1$ and $v_2$, MF will put both items close to $u_1$, but will not necessarily place $v_1$ and $v_2$ close to each other. That is to say, the seemingly obvious item--item similarity (between $v_1$ and $v_2$) is not guaranteed to be learned when the triangle inequality is violated~\cite{hsieh2017collaborative}.

To address the above limitation of MF--based recommendation, several metric learning approaches have been proposed
% for various recommendation tasks
~\cite{khoshneshin2010collaborative,chen2012playlist,feng2015personalized,hsieh2017collaborative}.
%	Although these approaches are targeted to solve different tasks,
They commonly project entities (users and items) into a low-dimensional metric space, i.e., Euclidean space, where the similarity between points is inversely proportional to the Euclidean distance that satisfies the triangle inequality.
Specifically, CML~\cite{hsieh2017collaborative} is the state-of-the-art 
metric learning--based recommendation method for implicit feedback (e.g., clicks or purchases);
%	metric learning method for implicit feedback-based recommendation 
it minimizes the distances between embeddings of a user and items that the user has interacted with, under the assumption that a user should be closer to the items the user likes than to those that the user does not.
%the positive interactions of user--item pairs of implicit feedback data (clicks, bookmarks, played).
In this way, these approaches not only expect to propagate positive user--item relationships to other unknown user--item pairs, but also to capture the similarity within user--user and item--item pairs by satisfying the triangle inequality.
%which pulls the positive pairs closer and pushes the other pairs relative further apart. By doing so, they aimed to learn a metric space where not only users' preferences on items, but also the user-user and item-item similarity are encoded.
%At the same time, because CML maintains the triangle inequality, the user-user and item- item similarity is also encoded in their euclidean distances in this joint space.

%Although previous metric learning approaches have shown their effectiveness by satisfying the triangle inequality, they suffer from an inherent limitation. i.e., each user is projected to a single point in the metric space. Such hard restrictions imposed on user embeddings preclude the previous methods from modeling various latent relationships of implicit user--item interactions. Note that a user’s implicit feedback on multiple items does not necessarily indicate his equal preference on these items, implying that every implicit user--item interaction encodes different latent relationship. Moreover, although a user may have various tastes on items

Although existing metric learning approaches have shown their effectiveness by satisfying the triangle inequality, they suffer from an inherent limitation, which is that \textit{each user is projected to a single point in the metric space}. Such a rigid restriction imposed on user embeddings makes it hard to properly model the {\textbf{intensity}} and the {\textbf{heterogeneity}} of user--item relationships in implicit feedback.
%However, existing metric learning approaches have an inherent limitation, that is, \emph{each user is projected to a single point in the metric space.} Such hard restriction imposed on user embeddings makes it hard to properly model the {\textbf{intensity}} and the {\textbf{heterogeneity}} of user--item relationships in implicit feedback.
%, which hinders further improvements.
%First, \textit{diverse} latent relationships of implicit user--item interactions cannot be properly modeled. 
More precisely, a user's implicit feedback  on multiple items does not necessarily indicate that he holds an equal preference for these items; rather, some of the items are more relevant to the user than others. 
%This implies that the \textit{intensity} of the user--item relationship embodied in every implicit user--item interaction differs from one another.
This implies that the \textit{intensity} of the user--item relationships embodied in implicit user--item interactions differ from one another.
%This implies that each user--item interaction embodies different \textit{strength} of user--item relationship. 
Moreover, a user may have a wide variety of tastes in different item categories, implying that 
the type of user--item relationship is \textit{heterogeneous} with regard to the user's tastes in various item categories\footnote{In Section~\ref{sec:Analysis}, we show the existence of the intensity and the heterogeneity of user--item relationships in implicit feedback.}.
%every implicit user--item interaction encodes \textit{heterogeneous} type of user--item relationship.
%\footnote{A user may be fond of multiple genres of movies, such as action and romance.}
%the types of user--item relationships 
%embodied in every implicit user--item interaction differs from one another.
%are \textit{heterogeneous} with regard to the item categories.
%diverse user--item relationship is embodied in every implicit user--item interaction.
%every implicit user--item interaction embodies \textit{diverse} latent relationship. 
However, it is by no means an easy task to project each user to a single point such that his intense and heterogeneous relationships with items are fully considered,
%\vspace{-1ex}
%the intensity and the heterogeneity of the user--item relationships are fully considered, 
and this phenomenon compounds as the number of users and items increases.

This paper presents a novel method called~\propose~to overcome the above limitation of existing metric learning approaches.
We achieve this by adopting the \textit{translation mechanism}, which has been proven to be effective for knowledge graph embedding~\cite{bordes2013translating,lin2015learning}, by which the relations between entities are interpreted as translation operations between them. 
In our work, we embed users and items as points in a low-dimensional metric space, and additionally introduce translation vectors to translate each user to multiple points.
Equipped with the user--item specific translation vectors, a user is translated toward his relevant items by considering his relationships with the items,
%, while being translated farther away from his non-relevant items.
%More precisely, each user is translated closer to the items relevant to him by considering his relationships with the items.
%Equipped with the user--item specific translation vectors, 
\vspace{-0.01ex}
which facilitates the modeling of the intensity and the heterogeneity of user--item relationships in implicit feedback that were overlooked by the previous metric learning approaches.
%users’ \textit{diverse} latent relationships regarding different items by translating a user to multiple points according to his relationship with items.
A further appeal of our proposed method is the ability to handle the \textit{complex} nature of CF where it is common for a user to interact with multiple items, i.e., one-to-many mapping.
%Whereas it is not feasible for the previous metric learning approaches to pull a user closer to all of the items because a user is projected to a single point, our proposed method alleviates this issue
Whereas it is not feasible for the previous metric learning approaches to pull a user closer to all of the items (one-to-many mapping) because a user is projected to a single point, our proposed method alleviates this issue
%if a user is projected to a single point, it is not feasible to pull a user close to all the items. On the other hand, if a user is translated to multiple points,
%because each translated user can be put close to the corresponding item.
because once a user is translated to multiple points, 
a one-to-many mapping can be deemed as multiple one-to-one mappings. 
%As a concrete example of the above, consider the following toy example.
As an illustration, consider the following toy example.
%regarding the user's relationships with items.
%which the latent relationship between user--item interactions is modeled. 
%	Our core idea is that, since the translation vectors are user--item specific, we are now able to model diverse and complex latent relationships from implicit user--item interactions, and thus naturally have points spread out in the latent space unlike CML.
%\begin{newthm*}
%	\vspace{-1ex}
%	\label{toyex}
%	\leftskip=0.1cm
%	\rightskip=0.1cm
%	\normalfont
%	\noindent
%	Figure~\ref{fig:explanation} shows a toy example that visually illustrates the benefit of adopting the \textit{translation mechanism}. While CML tries to find a single point for a user that minimizes the distance between the user and his relevant items,~\propose~introduces the user--item specific translation vectors to translate the user to multiple points regarding the intensity and the heterogeneity of user--item relationships. Note that the more intense the user--item relationship, the closer the user is expected to be translated to the item. \textcolor{red}{TODO: One more sentence regarding the heterogeneity of the user--item relationship after doing the cosine similarity experiments.}
%	\vspace{-1ex}
%\end{newthm*}	

\begin{figure}
	\centering
	\includegraphics[width=0.45\textwidth]{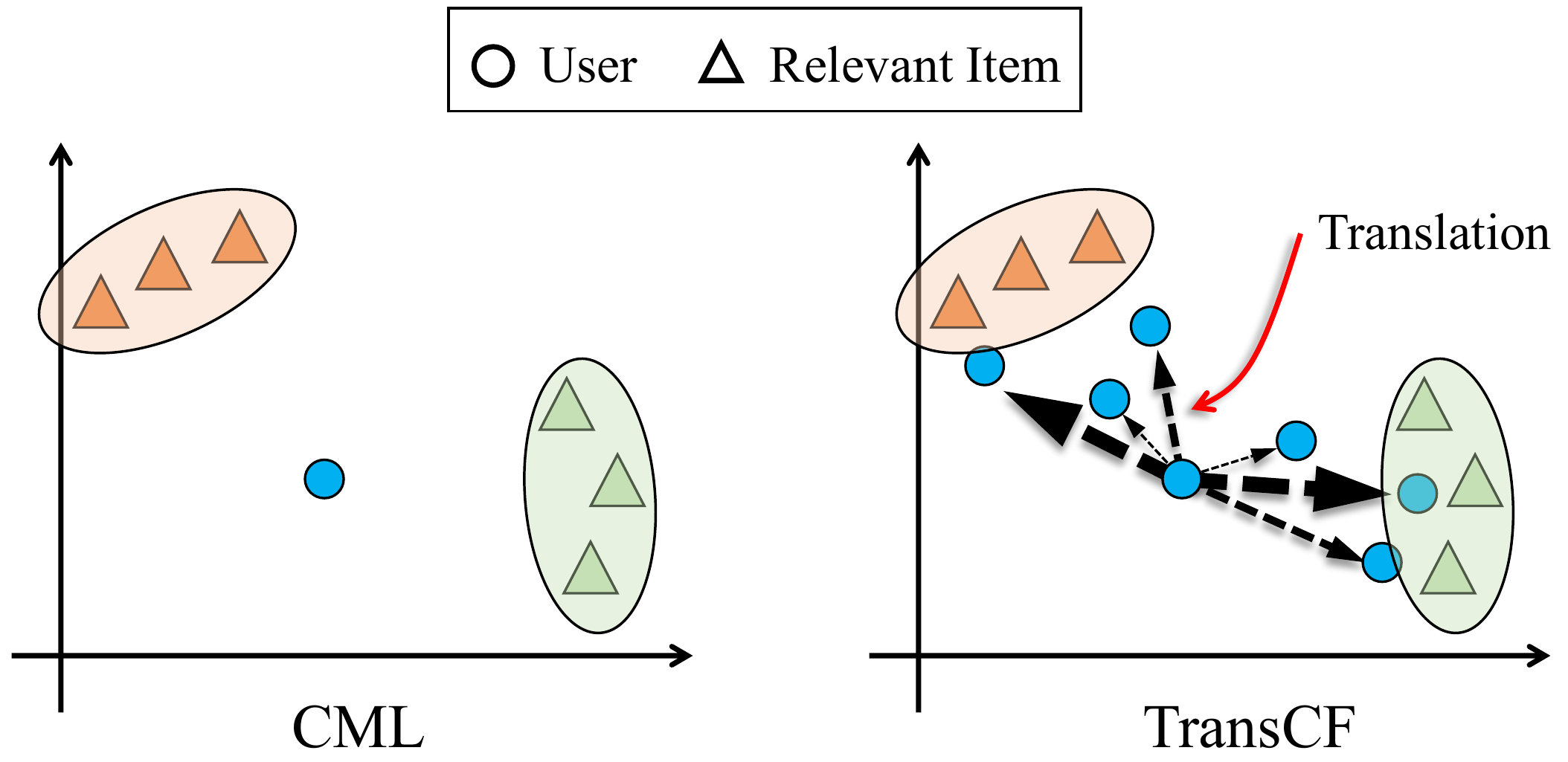}
	\caption{
	Toy example illustrating the benefit of the user--item specific translation vectors. Items are assumed to be grouped according to their categories.
%	The thickness of the vectors represents the {intensity} of user--item relationships, and items are grouped by their categories to show that the user--item relationships are {heterogeneous} regarding the users' tastes in various item categories.
	}
	\label{fig:explanation}
	\vspace{-2ex}
\end{figure}

\smallskip
\begin{adjustwidth}{0.1cm}{0.1cm}
	\noindent\textbf{\underline{Toy Example.}}	In Figure~\ref{fig:explanation}, whereas CML tries to find a single point for a user that minimizes the distances between the user and his relevant items,~\propose~introduces user--item specific translation vectors to translate the user to multiple points according to the intensity and the heterogeneity of the user--item relationships. The more intense (thickness of the vectors) the user--item relationship, the closer the user is expected to be translated to the item. 
	Besides, the direction of the vectors and the angles between them reflect the heterogeneity of the relationship with regard to the user's tastes in various item categories.
	%\noindent\textbf{\underline{Toy Example.}}	Figure~\ref{fig:explanation} shows a toy example that  illustrates the benefit of adopting the \textit{translation mechanism}. While CML tries to find a single point for a user that minimizes the distance between the user and his relevant items,~\propose~introduces the user--item specific translation vectors to translate the user to multiple points regarding the intensity and the heterogeneity of user--item relationships. Note that the more intense the user--item relationship, the closer the user is expected to be translated to the item, and the direction of the vectors reflect the heterogeneity of the relationship regarding the user's tastes in various item categories.
	%\textcolor{red}{TODO: One more sentence regarding the heterogeneity of the user--item relationship after doing the cosine similarity experiments.}
\end{adjustwidth}
\smallskip
%	In this paper, we propose a novel approach to overcoming the limitations of previous metric learning approaches by adopting the \textit{translational concept}, which is widely used for knowledge graph embedding ~\cite{bordes2013translating}.
%	We model diverse and complex relationships between user--item interactions by adopting the \textit{translational concept}, which has been proven to be effective for knowledge graph embedding ~\cite{bordes2013translating}.	
%	 the latent relationship between user--item interactions into low-dimensional relation embeddings 
%	Our work is inspired by the \textit{translational concept}, which has been proven to be effective for knowledge graph embedding ~\cite{bordes2013translating}, where the relations between entities are interpreted as translation operations between them.
%	To be precise, we embed user and items as points in a low-dimensional metric space, and connect them by translation vectors in which the latent relationship between user--item interactions is modeled.
%We introduce relation embeddings to distinguish different latent relationship and the complex nature of user--item interactions.
%in which the latent relationship between user--item interactions is modeled. 
%%%%% which is,,,,,,simple relation is modeled
%Precisely, we define the relationship between user and item embedding through relation embedding, where an item embedding vector is approximated by the sum of user and relation embedding vectors.

However, directly applying the translation mechanism into a general recommendation framework for implicit feedback is not viable because \textit{user--item interactions in the implicit feedback dataset are not labeled}, unlike in knowledge graphs where each relation between entities is given a label, such as {``was\_born\_in''} or {``performed\_in''}.
%\footnote{In knowledge graphs, each relation between entities is given a label, such as \textit{``was\_born\_in''} or \textit{``performed\_in''}.}. 
%For this reason, CML simply treats every implicit user--item interaction equally, and thus inadvertently fails to model their diverse and complex relationships as mentioned above.
%which eventually hinders further improvements. 
Therefore, the key for successfully adopting the translation mechanism in our framework boils down to
%to successfully adopting the translation mechanism to our framework is 
defining labels for implicit user--item interactions, and materializing them into the user--item specific translation vectors.
%	While it is possible to introduce a new parameter for each translation vector corresponding to a user--item pair, we note that it is not only prone to overfitting, but also the collaborative information cannot be explicitly modeled.
%To construct the translation vectors, 
Inspired by the highly effective neighborhood--based CF algorithms~\cite{koren2008factorization,kabbur2013fism,wu2016collaborative,ning2011slim,desrosiers2011comprehensive,sarwar2001item}, whose assumptions are that similar users display similar item preferences and that similar items are consumed by similar users, we employ the neighborhood information of users and items to construct the translation vectors (\textbf{Section~\ref{met:Neighbor}}).
%	We note that while these vectors can be flexibly constructed as long as the user--item interactions are properly modeled, here we construct them by employing the neighborhood information of users and items, which has been shown to be highly effective for recommendation~\cite{koren2008factorization,kabbur2013fism,wu2016collaborative}.
%By doing so, 
Combined with a regularizer, specifically tailored to~\propose, that guides each user/item to its neighbors (\textbf{Section~\ref{subsec:reg_nbr}}),~\propose~explicitly models the neighborhood information, in contrast to CML, which expects to achieve it implicitly by satisfying the triangle inequality. 
It is worth noting that {the user--item specific translation vectors are constructed without introducing any new parameters}, which prevents our proposed method from overfitting.
Furthermore, we introduce a second regularizer to further improve the accuracy of~\propose~in case the user--item relationships become more complex (\textbf{Section~\ref{subsec:reg_dist}}). 
%Our extensive experiments on four real-world datasets show that~\propose~significantly outperforms the state-of-the-art recommendation methods for implicit feedback by up to 16\% in terms of MRR@10. In addition, we perform various experiments to qualitatively verify the benefit of adopting the translation mechanism to implicit feedback based recommendations.
Our extensive experiments on seven real-world datasets (\textbf{Section~\ref{exp:quant}}) show that~\propose~considerably outperforms the state-of-the-art recommendation methods for implicit feedback by up to 17\% in terms of hit ratio. We also perform various experiments to qualitatively ascertain that~\propose~indeed benefit from modeling the intensity and the heterogeneity of user--item relationships in implicit feedback as illustrated in Figure~\ref{fig:explanation} (\textbf{Section~\ref{exp:qual}}).  The source code is available at https://github.com/pcy1302/TransCF.
%the benefit of adopting the translation mechanism for implicit feedback-based recommendations 
%The source code of our method is accessible at ``http://'' for reproducibility.

%	We postulate that defining labels on user--item interactions is the key to adopting the translation mechanism into implicit feedback-based recommendation task, and that the translation vectors can be formed flexibly as long as the user--item interactions can be modeled. 

\captionsetup[sub]{textfont=normalfont}
\begin{figure}[h]
	\centering
	\begin{subfigure}{\linewidth}%
		\centering\captionsetup{width=\linewidth}%
		\includegraphics[width=0.92\linewidth]{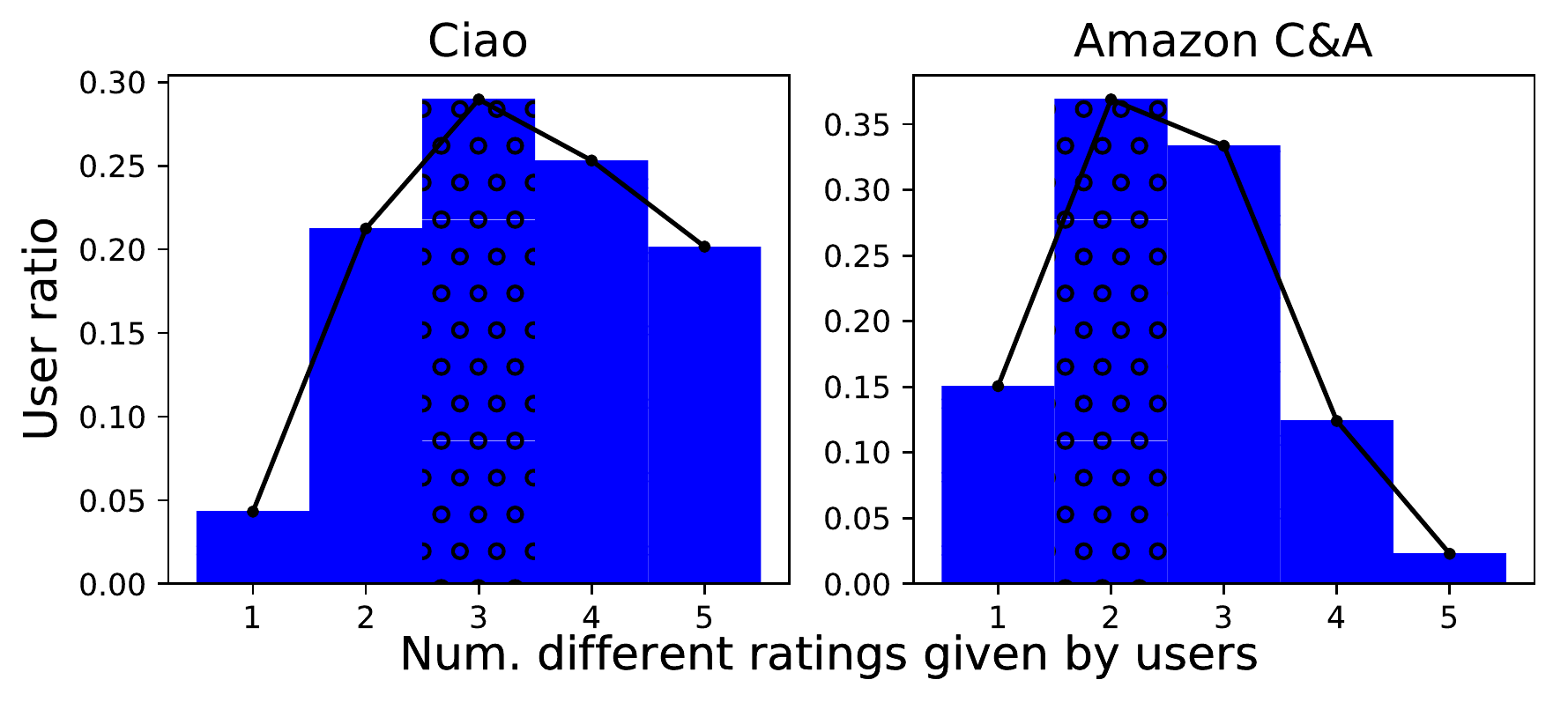}%
		%		\vspace{-1.5ex}
		\caption{Distribution of the number of different ratings given by users.}
		\label{fig:analysis:rating}
	\end{subfigure}
	\\
	\begin{subfigure}{\linewidth}%		
		\centering\captionsetup{width=\linewidth}%
		\includegraphics[width=0.92\linewidth]{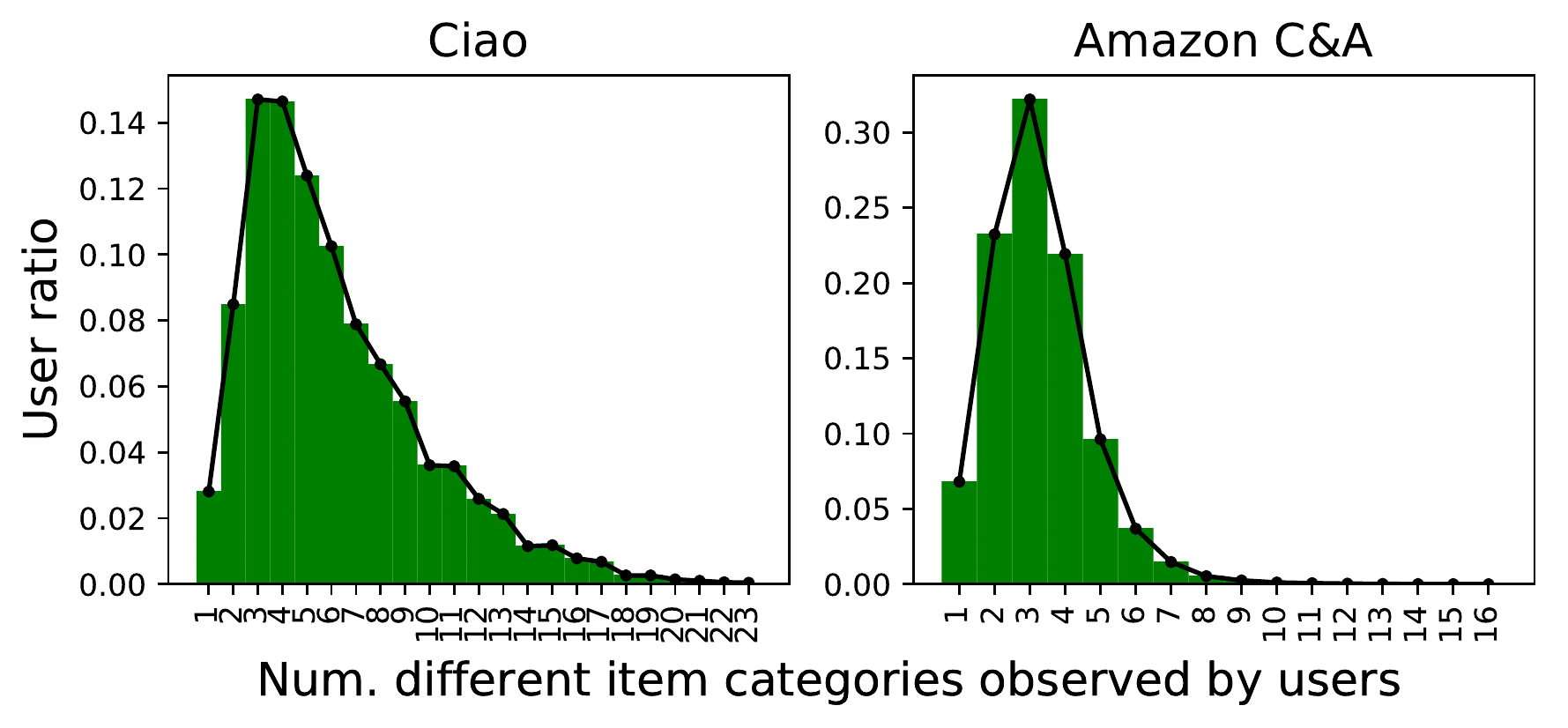}%
		%		\vspace{-1.5ex}
		\caption{Dist. of the num. different item categories observed by users.}
		\label{fig:analysis:category}
	\end{subfigure}
	%	\vspace{-2.5ex}
	\caption{Data analysis on Ciao and Amazon C\&A datasets.}
	\label{fig:analysis}
	\vspace{-1ex}
\end{figure}

\section{Data Analysis: \small Intensity and Heterogeneity}
\label{sec:Analysis}
In this section, we perform data analyses on Ciao~\cite{tang-etal12a} and Amazon C\&A~\cite{he2016ups} datasets to provide evidences of the existence of the \textit{intensity} and the \textit{heterogeneity} of user--item relationships in implicit feedback.
The ratings in both datasets are integers from 1 to 5, and the numbers of unique item categories are 28 and 45, respectively. In our experiments, we regard every user--item interaction with a rating as an implicit feedback record\footnote{It is unavoidable to use explicit feedback datasets as it is not feasible to judge the intensity of user--item relationships from implicit feedback.}.
Figure~\ref{fig:analysis} shows the distribution of the number of different ratings and different item categories given and observed by users.
In Figure~\ref{fig:analysis:rating}, we observe that most users (95\% in Ciao and 85\% in Amazon C\&A) give two or more different ratings. This implies that implicit user--item interactions do indeed encode different \textbf{intensities} of user--item relationships, assuming that the rating is a proxy for the intensity; a higher rating implies higher intensity. Moreover, in Figure~\ref{fig:analysis:category} we observe that only a few users (3\% in Ciao and 6\% in Amazon C\&A) interact with a single category, whereas the vast majority of users interact with two or more different item categories. This implies that users have diverse tastes in various item categories, and thus the type of user--item relationship is \textbf{heterogeneous} with regard to the users' tastes in various item
categories. 

In Section~\ref{sec:int_het}, we show by our experiments that~\propose~can inversely infer knowledge regarding the intensity and the heterogeneity from the implicit user--item interactions.
%the rating and the item category information from the implicit user--item interactions.

%\vspace{-0.5ex}
\section{Proposed Method:~\textsf{T}\lowercase{\textsf{rans}}\textsf{CF}}
In this section, we give details of our proposed method, which adopts the translation mechanism for modeling the intensity and the heterogeneity of user--item relationships in implicit feedback.

%\vspace{-1ex}
\subsection{Problem Formulation}
In this work, we focus on recommendations for implicit feedback. 
%We first introduce notations used throughout this paper. 
Let $\mathcal{U}$ and $\mathcal{I}$ denote a set of users and a set of items, respectively.
%We have a set of users and a set of items, denoted as $\mathcal{U}$ and $\mathcal{I}$, respectively.
%	Let $\mathcal{U}= \{ {u_1},{u_2},...,{u_n}\}$ be the set of users and $\mathcal{I} = {\{i_1},{i_2},...,{i_m}\}$ be the set of items, where $n$ and $m$ are the number of users and the number of items, respectively. 
%	The implicit user--item interactions are represented by the rating matrix $\mathcal{R}=[\mathcal{R}_{ui}]_{n\times m}$, where $\mathcal{R}_{ui}=1$ if user $u$ interacted with item $i$, and 0 otherwise. 
$\mathcal{N}^\mathcal{I}_u$ denotes the set of items that user $u$ has previously interacted with.
%, and $\mathcal{N}^\mathcal{U}_i$ denotes the set of users that previously interacted with item $i$.
Given implicit user--item interactions (e.g., bookmarks and purchase history), our goal is to recommend items $i\in\mathcal{I}\backslash \mathcal{N}^\mathcal{I}_u$ to each user $u\in\mathcal{U}$. Note that we use neither the rating information nor any auxiliary information regarding users and items (e.g., user social network or item category). 
%The notations are summarized in Table~\ref{tab:notation}.
%\begin{figure}[t]
%	\centering
%	\subfigure[Distribution of the number of different ratings given by users.]{\label{fig:analysis:rating}\includegraphics[width=0.7\linewidth]{./figure/rating.pdf}}
%	\\[-1ex]
%	\subfigure[Distribution of the number of different item categories observed by users.]{\label{fig:analysis:category}\includegraphics[width=0.7\linewidth]{./figure/cat.pdf}}
%	\vspace{-3ex}
%	\caption{Data analysis on Ciao and Amazon C\&A datasets.}
%	\label{fig:analysis}
%	\vspace{-4ex}
%\end{figure}

%\vspace{-1ex}
\subsection{Scoring Function}
Recall that the objectives of this work are 1) to address the limitation of the inner product as a scoring function, which violates the triangle inequality, and 2) to model the intensity and the heterogeneity of user--item relationships in implicit feedback. To this end, we adopt Euclidean distance as our distance metric to satisfy the triangle inequality, and we introduce translation vectors to model the intensity and the heterogeneity of implicit user--item interactions. Given $K$-dimensional embedding vectors $\bm{\alpha}_u\in\mathbb{R}^K$ for user $u$, $\bm{\beta}_i\in\mathbb{R}^K$ for item $i$, and $\bm{r}_{ui}\in\mathbb{R}^K$ for the translation of user $u$ with regard to item $i$, the similarity score $s(u,i)$ between user $u$ and item $i$ is:
% computed:
\begin{equation}
%	\small
	\label{eqn:score}
	s(u,i) = -\norm{\bm{\alpha}_u + \bm{r}_{ui} - {\bm{\beta}_i}}_2^2
\end{equation}	
where a higher similarity score $s(u,i)$ implies a higher probability that user $u$ will like item $i$.
%a higher similarity score between user $u$ and item $i$, and the higher the similarity score the more likely for the user $u$ will like item $i$. 
In other words, the similarity score between user $u$ and item $i$ is computed by the distance between the translated embedding vector of user $u$, given by ($\bm{\alpha}_u+\bm{r}_{ui}$), and the embedding vector $\bm{\beta}_i$ of item $i$.
%As will be explained in Section~\ref{met:Neighbor}, the translation vector $\bm{r}_{ui}$ is user--item specific. 
%That is, every implicit user--item interaction has a different translation vector, and each user is translated with respect to multiple items regarding his relationships with the items.

\medskip
\noindent\textbf{Optimization Objective. }
Given the scoring function $s(u,i)$ as in Equation~\ref{eqn:score}, we minimize a popular margin--based pairwise ranking criterion~\cite{hsieh2017collaborative,bordes2013translating,lin2015learning}, i.e., hinge loss, as follows:
\begin{equation}
%	\small
	\label{eqn:obj}
	\mathcal{L}(\Theta) = \sum\limits_{u \in \mathcal{U}} {\sum\limits_{i \in {\mathcal{N}^\mathcal{I}_u}} {\sum\limits_{j \notin {\mathcal{N}^\mathcal{I}_u}} { [{\gamma  - s(u,i) + s(u,j)} ]_{+} } } }
\end{equation}
where $[x]_+ = max(x,0)$, and $\gamma$ is the margin. Our objective is to ensure that the similarity score of an observed (positive) user--item pair $(u,i)$ is higher than that of an unobserved (negative) pair $(u,j)$ by a margin of at least $\gamma$.
By doing so, we aim to translate each user $u$ closer to his relevant item $i$ while translating him farther away from his non-relevant item $j$.
One thing to note is that the translation vector $\bm{r}_{ui}$ is user--item specific: \textit{it translates user $u$ with respect to item $i$ according to the user's specific relationship with the item $i$.}
%implying that it encodes the latent relationship between user $u$ and item $i$, which is specific to them.
This property enables~\propose~not only to capture the intensity and the heterogeneity of user--item relationships in implicit feedback, but also to handle the complex nature of CF.
%\begin{table}[]
%	\centering
%	\small
%	\captionsetup{font=normal}
%	\caption{Notations.}
%	\vspace{-2ex}
%	\label{tab:notation}
%	\begin{tabular}{l|l}
%		\specialrule{.1em}{.1em}{.1em} 
%		Symbol & \multicolumn{1}{c}{Description} \\ \specialrule{.1em}{.1em}{.1em}
%		$\mathcal{U}, \mathcal{I}$& Set of Users, Set of Items                                 \\ 		
%		$n, m$& Number of users and items                                 \\ 
%		$\mathcal{N}^\mathcal{I}_u$ & Items that user $u$ previously interacted with \\ 
%		$\mathcal{N}^\mathcal{U}_i$ & Users that previously interacted with item $i$ \\
%		${s(u,i)}$& The similarity score between user $u$ and item $i$ \\
%		$K$& Number of latent dimensions                                 \\ 
%		$\bm{\alpha}_u \in \mathbb{R}^{K}$& Embedding vector of user $u$                                 \\ 
%		$\bm{\beta}_i \in \mathbb{R}^{K}$& Embedding vector of item $i$						\\ 		
%		$\bm{r}_{ui} \in \mathbb{R}^{K}$& Translation embedding of user $u$ regarding item $i$						\\ 		
%		$\bm{\alpha}_u^{\mathrm{nbr}} \in \mathbb{R}^{K}$& Neighborhood embedding vector of user $u$                                 \\ 
%		$\bm{\beta}_i^{\mathrm{nbr}} \in \mathbb{R}^{K}$& Neighborhood embedding of item $i$						\\ 		
%		${\lambda_{\mathrm{nbr}}}$& The strength of the neighborhood regularizer \\ 
%		${\lambda_{\mathrm{dist}}}$& The strength of the distance regularizer \\ 		
%		${\eta}$& Learning rate \\
%		${\gamma}$& Margin \\
%		\specialrule{.1em}{.1em}{.1em}
%	\end{tabular}
%\end{table}
\subsection{Modeling Relations: Neighborhood Information}
\label{met:Neighbor}
As mentioned previously, the key for successfully adopting the translation mechanism in our framework is modeling the translation vectors so that they reflect the
intensity and the heterogeneity of user--item relationships in implicit feedback.
%reflect the diverse and complex nature of user--item interactions. 
The translation vectors can be flexibly constructed as long as the user--item interactions are properly modeled.
Although it is possible to introduce a new parameter for each translation vector corresponding to every user--item pair, we note that not only is this approach prone to overfitting owing to the large number of parameters, but also it does not allow the collaborative information to be explicitly modeled as no parameters are shared among users and among items.
Therefore, in this work we employ the neighborhood information of users and items to construct the translation vectors without introducing any new parameters.
Neighborhood information, which has been shown to be highly effective for recommendation~\cite{wu2016collaborative,ning2011slim,desrosiers2011comprehensive,sarwar2001item}, implies that users are represented through the items that they prefer~\cite{koren2008factorization,kabbur2013fism}, and items are represented through the users that prefer them.
More precisely, considering that user embedding vectors reveal users' tastes, each item can be regarded as the average tastes of the users that have interacted with that item:
\begin{equation}
%	\small
	\label{eqn:itemnbr}
	\bm{\beta}_i^{\mathrm{nbr}}={\frac{1}{{|\mathcal{N}_i^\mathcal{U}|}}\sum\limits_{k \in \mathcal{N}_i^\mathcal{U}} {{\bm{\alpha}_k}} }
\end{equation}
where $\bm{\beta}_i^{\mathrm{nbr}}\in\mathbb{R}^K$ denotes the neighborhood embedding of item $i$, and $\mathcal{N}^\mathcal{U}_i$ denotes the set of users that have previously interacted with item $i$. 
Likewise, considering that item embedding vectors capture the prominent features of items, e.g., the item category, each user's taste can be represented by the average features of the items that the user has interacted with: 
\begin{equation}
%	\small
	\label{eqn:usernbr}
	\bm{\alpha}_u^{\mathrm{nbr}}={\frac{1}{{|\mathcal{N}_u^\mathcal{I}|}}\sum\limits_{k \in \mathcal{N}_u^\mathcal{I}} {{\bm{\beta} _k}} }
\end{equation}
where $\bm{\alpha}_u^{\mathrm{nbr}}\in\mathbb{R}^K$ denotes the neighborhood embedding of user $u$, and $\mathcal{N}^\mathcal{I}_u$ denotes the set of items that user $u$ has previously interacted with.
Given the respective representations of an item and a user from the neighborhood perspective as in Equations~\ref{eqn:itemnbr} and~\ref{eqn:usernbr}, we use them to model the user--item interactions as follows:
\begin{equation*}
%	\small
	%	\bm{r}_{ui} = \bm{\alpha}_u^{\mathrm{nbr}} \odot  \bm{\beta}_i^{\mathrm{nbr}}
	\bm{r}_{ui} = f(\bm{\alpha}_u^{\mathrm{nbr}},  \bm{\beta}_i^{\mathrm{nbr}})
\end{equation*}
where $f(x,y)$ denotes a function to model the interaction between the two input vectors $x$ and $y$. Note that although we could employ various functions such as a multi-layer perceptron,
%instead of element-wise products to model the interactions in a more complex way, however
it turns out that a simple element-wise vector product provides the highest accuracy.
%Note that we exclude the user and item itself into our neighborhood computations, because 
% Did not include the user and item itself. Known to be better in FISM. Excluding the current item in the similarity computation

\begin{figure}
	\centering
	\includegraphics[width=0.4\textwidth]{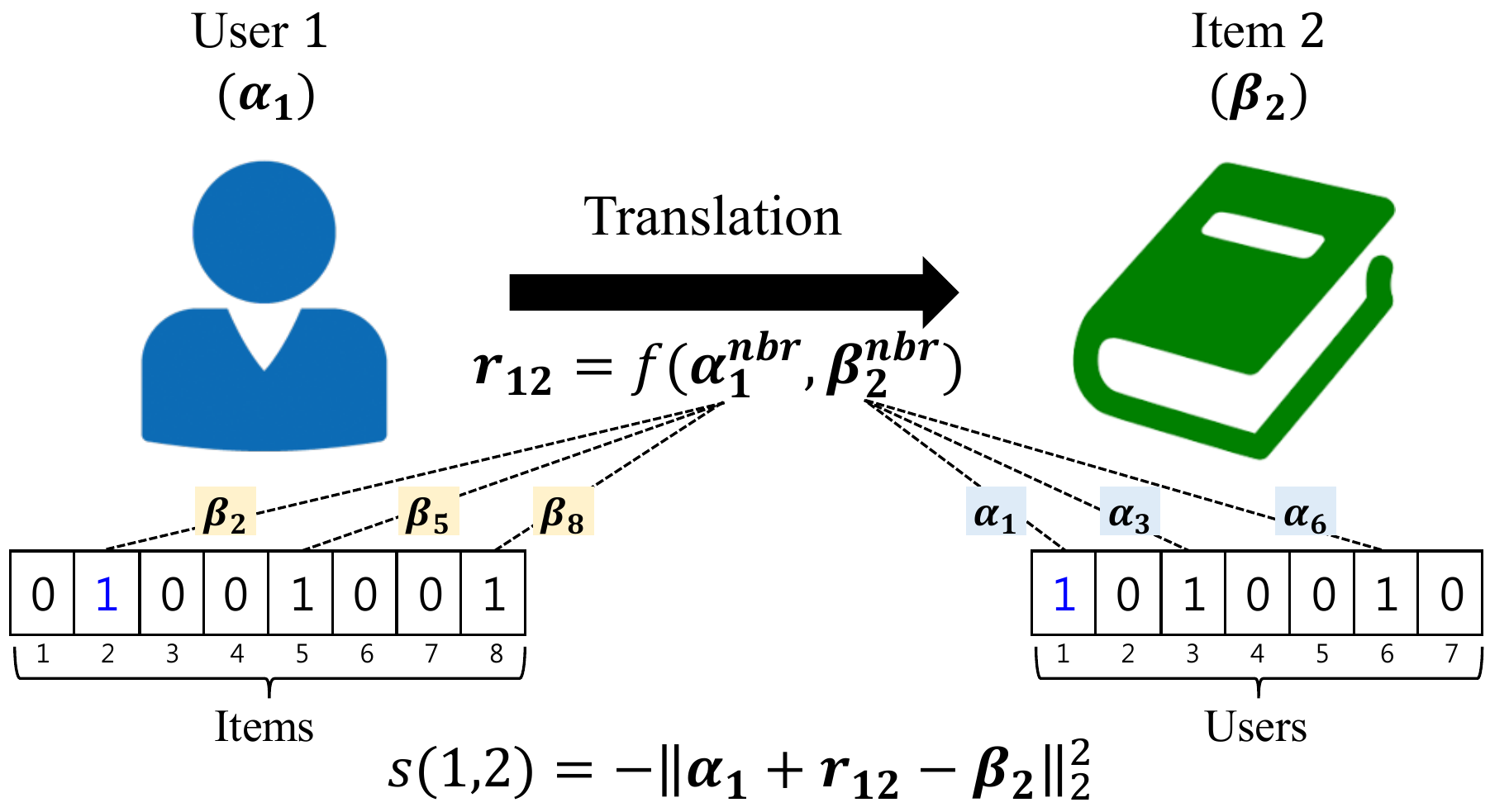}
%	\vspace{-2ex}
	\caption{An overview of~\propose. }
	\label{fig:overview}
	\vspace{-3ex}
\end{figure}

%In Figure~\ref{fig:overview}, we illustrate an example working scenario of~\propose. 
Figure~\ref{fig:overview} illustrates a working example of~\propose.
Given that user $1$ has interacted with items $\{2,5,8\}$ and item $2$ has been interacted with by users $\{1,3,6\}$, 
%Then, given the embedding vector $\bm{\alpha}_1$ of user 1, 
we obtain the neighborhood embedding vector $\bm{\alpha}_1^{\mathrm{nbr}}$ of user $1$ from the set of items $\{2,5,8\}$, and the neighborhood embedding vector $\bm{\beta}_2^{\mathrm{nbr}}$ of item 2 from the set of users $\{1,3,6\}$. Then, using these neighborhood embedding vectors $\bm{\alpha}_1^{\mathrm{nbr}}$ and $\bm{\beta}_2^{\mathrm{nbr}}$, we construct the translation embedding vector $\bm{r}_{12}=f(\bm{\alpha}_1^{\mathrm{nbr}},  \bm{\beta}_2^{\mathrm{nbr}})$, which is eventually used for computing the similarity score $s(1,2)=-\norm{\bm{\alpha}_1+\bm{r}_{12}-\bm{\beta}_2}_2^2$.
%combined with $\bm{\alpha}_1$ and $\bm{\beta}_2$.

\smallskip
%	\paragraph{Discussion}
%\noindent\textbf{Discussion.}

%\vspace{-1.5ex}
\subsubsection{\textbf{Discussion}}
\label{sec:discussion}
If users' social network information or auxiliary information related to items is given, it would be more intuitive to represent a user by his friends' tastes, and an item by its semantically related items, e.g., items that belong to the same category. However, such side information on users and items is not always provided in practice, and thus we do not consider any side information in our current work. 
Additionally, instead of simply averaging the embeddings of neighbors as in Equations~\ref{eqn:itemnbr} and~\ref{eqn:usernbr}, we could expect further improvements by applying the attention mechanism~\cite{bahdanau2014neural}, whereby we 
%compute the weighted sum of neighbors. 
give higher weights to more influential neighbors.
Lastly, we could try swapping the role of users and items such that we conversely translate items towards users.
However, since the above extensions are straightforward and our main focus is to incorporate the translation mechanism into recommender systems, we leave these for future studies.

%\subsection{Optimization Objective}
%Given the scoring function $s(u,i)$ as in Equation~\ref{eqn:score}, we minimize a margin-based pairwise ranking criterion, i.e., hinge loss, as follows:
%\begin{equation}
%\label{eqn:obj}
%\mathcal{L}(\Theta) = \sum\limits_{u \in \mathcal{U}} {\sum\limits_{i \in {\mathcal{N}^\mathcal{I}_u}} {\sum\limits_{j \notin {\mathcal{N}^\mathcal{I}_u}} { [{0,\gamma  - s(u,i) + s(u,j)} ]_{+} } } }
%\end{equation}
%where $[x]_+ = max(x,0)$ is the hinge loss, $\gamma$ is the margin. That is, we aim to ensure that the similarity score of an observed user--item pair $(u,i)$ is higher than that of an unobserved pair $(u,j)$ at least by margin $\gamma$. 
%For each user $u$, we sample $P$ pairs of $(i,j)$ from the training set $\mathcal{O}=\{\mathcal{O}_u | u\in|\mathcal{U}|\}$ where
%$	\mathcal{O}_u=\{(i,j)|i \in \mathcal{N}^\mathcal{I}_u \wedge  j \in \mathcal{I} \char`\\ \mathcal{N}^\mathcal{I}_u\}$, $\mathcal{O}_u$ denoting the training set for user $u$.

%\vspace{-2ex}
\subsection{Model Regularization}
\vspace{-0.5ex}
In this section, we introduce two regularizers that are tailored to~\propose.
%beneficial for enhancing the model performance. 
We later show in our experiments (Section~\ref{sec:exp_reg}) that these regularizers are indeed beneficial.

\medskip
\subsubsection{\textbf{Regularizer 1 $-$ Neighborhood Regularizer}}
\label{subsec:reg_nbr}
%\noindent{\textbf{Regularizer 1 $-$ Neighborhood Regularizer. }}
In Section~\ref{met:Neighbor}, we explained how we reflect the neighborhood information of users and items into our translation vectors, under the assumption that users and items can be represented by their neighbors. In the case of users, we implicitly assumed that $\bm{\alpha}_u$ can be represented by $\bm{\alpha}^{\mathrm{nbr}}_u$ (Equation~\ref{eqn:usernbr}), and likewise for items, that $\bm{\beta}_i$ can be represented by $\bm{\beta}^{nbr}_i$ (Equation~\ref{eqn:itemnbr}).
% in Equation~\ref{eqn:itemnbr}.
However, to ensure that the neighborhood information is explicitly incorporated into our model, we need to guide $\bm{\alpha}_u$ to be close to $\bm{\alpha}^{\mathrm{nbr}}_u$, and $\bm{\beta}_i$ to be close to $\bm{\beta}^{\mathrm{nbr}}_i$.
%However, by guiding $\bm{\alpha}_u$ to be close to $\bm{\alpha}^{\mathrm{nbr}}_u$ and $\bm{\beta}_i$ to be close to $\bm{\beta}^{\mathrm{nbr}}_i$, we can be ascertained that the neighborhood information will be explicitly reflected into our model.
%	We postulate that if we have a label for which these neighborhood models can be ,
%	While we implicitly encoded the neighborhood information into our model, the neighborhood model guide	
%In order to get more robust representation of users/items and their neighbors, 
% label of neighbor guide
To this end, we introduce a regularizer named $re{g_{\mathrm{nbr}}}(\Theta)$ that minimizes the distance between users/items and their neighbors:
\begin{equation}
	\small
	\label{eqn:reg_nbr}
	\begin{split}
	re{g_{\mathrm{nbr}}}(\Theta ) &= {\sum\limits_{u \in \mathcal{U}} {\left( {{\bm{\alpha} _u} - \frac{1}{{|\mathcal{N}_u^\mathcal{I}|}}\sum\limits_{k \in \mathcal{N}_u^\mathcal{I}} {{\bm{\beta} _k}} } \right)} ^2} \\
	&+ {\sum\limits_{i \in \mathcal{I}} {\left( {{\bm{\beta} _i} - \frac{1}{{|\mathcal{N}_i^\mathcal{U}|}}\sum\limits_{k \in \mathcal{N}_i^\mathcal{U}} {{\bm{\alpha} _k}} } \right)} ^2}
	\end{split}
\end{equation}
Through Equation~\ref{eqn:reg_nbr}, we aim to explicitly inject the neighborhood information into the users and items, leading to more robust representations of users/items and their neighbors.

\medskip
%\noindent{\textbf{Regularizer 2 $-$ Distance Regularizer. }}
\subsubsection{\textbf{Regularizer 2 $-$ Distance Regularizer}}
\label{subsec:reg_dist}
The objective function shown in Equation~\ref{eqn:obj} trains the user and item embeddings so that given a positive user--item interaction $(u,i)$, the item embedding $\bm{\beta}_i$ is the nearest neighbor of the user embedding $\bm{\alpha}_u$ translated by the translation vector $\bm{r}_{ui}$; i.e., $\bm{\alpha}_u + \bm{r}_{ui} \approx \bm{\beta}_i$. 
In other words, the objective function (Equation~\ref{eqn:obj}) expects that the positive item $i$ will be pulled toward user $u$ by pushing the negative item $j$ away from user $u$, which is in fact what is done in CML.
However, the relations become more complex as the number of user--item interactions grows, and it is crucial to guarantee that the actual distance between them is small. Therefore, we introduce a second regularizer named $re{g_{\mathrm{dist}}}(\Theta)$ to explicitly pull the item embedding closer to the translated user embedding as follows:
\begin{equation}
	\small
	\label{eqn:reg_dist}
	re{g_{\mathrm{dist}}}(\Theta ) = \sum\limits_{u \in \mathcal{U}} {\sum\limits_{i \in {\mathcal{N}^\mathcal{I}_u}} {-s(u,i)} }  = \sum\limits_{u \in \mathcal{U}} {\sum\limits_{i \in {\mathcal{N}^\mathcal{I}_u}} {\norm{\bm{\alpha}_u + \bm{r}_{ui} - \bm{\beta}_i}_2^2} } 
\end{equation}
Notably, $re{g_{\mathrm{dist}}}(\Theta)$ is equivalent to the loss $\mathcal{L}_{\mathrm{pull}}$, which is introduced in the paper that proposed CML~\cite{hsieh2017collaborative}. 
However, CML does not employ $\mathcal{L}_{\mathrm{pull}}$ because \textit{``an item can be liked by many users and it is not feasible to pull it closer to all of them''} as the authors mentioned in Section 3.1 of their paper~\cite{hsieh2017collaborative}. This infeasibility is essentially caused by the fact that each user is projected to a single point,
%We postulate that this happens because each user is projected to a single point.
%if an item is liked by many users, CML cannot pull the item close to all of the users, as mentioned in their paper~\cite{hsieh2017collaborative}, because each user is projected to a single point.
and we argue that translating a user to multiple points by introducing the user--item specific translation vectors as in our method makes it feasible to pull the item closer to all of the translated users, allowing~\propose~to employ $re{g_{\mathrm{dist}}}(\Theta)$.
%However, introducing the user--item specific translation vector as in our method makes it feasible to pull the item close to all of the translated users.
%	However, introducing translation vector as in our method makes this feasible, because each user is translated with respect to the item by using the user--item specific translation vector, which makes it feasible to pull the item close to all of the translated users.
%user--item specific translation vector helps to avoid pushing entities to a single point.

\medskip
\noindent{\textbf{Final Objective Function.}}
Given the margin--based pairwise ranking function (Equation~\ref{eqn:obj}) and the two regularizers (Equations~\ref{eqn:reg_nbr} and~\ref{eqn:reg_dist}), our final objective function $\mathcal{J}(\Theta)$ to minimize is as follows:
\begin{align*}
%	\small
	\mathcal{J}(\Theta) = \left( \mathcal{L}(\Theta) + \lambda_{\mathrm{nbr}}\cdot re{g_{\mathrm{nbr}}}(\Theta ) + \lambda_{\mathrm{dist}}\cdot re{g_{\mathrm{dist}}}(\Theta ) \right) 
\end{align*}
where $\lambda_{\mathrm{nbr}}$ and $\lambda_{\mathrm{dist}}$ are regularization coefficients for the neighborhood regularizer and the distance regularizer, respectively. We compute the gradient for parameters in $\Theta=\{\bm{\alpha}_u,\bm{\beta}_i\}$, and update them by using mini-batch stochastic gradient descent (SGD) with learning rate $\eta$ as follows:
%\begin{align*}
$	\Theta \leftarrow \Theta - \eta \times \frac{\partial\mathcal{J}(\Theta)}{\partial\Theta}$.
%\end{align*}
As our focus is to verify the benefit of translation mechanism for recommendation, we do not adopt advanced negative sampling techniques~\cite{he2016fast,pan2008one,rendle2014improving,weston2011wsabie} for the model training.
Instead, for each user $u\in\mathcal{U}$, we randomly generate 100 samples of $\{(i,j)|i\in{\mathcal{N}^\mathcal{I}_u} \cap j\notin{\mathcal{N}^\mathcal{I}_u}\}$ in every epoch. 
Moreover, as done in CML, we apply regularizations on $\bm{\alpha}_*$ and $\bm{\beta}_*$ after each epoch by bounding them within a unit sphere to mitigate `curse of dimensionality' issue~\cite{bordes2013translating,lin2015learning}: $\norm{\bm{\alpha}_*}^2 \leq 1$ and $\norm{\bm{\beta}_*}^2 \leq 1$, which is achieved by $\bm{\alpha}_*\leftarrow\bm{\alpha}_*/\max(1,\norm{\bm{\alpha}_*}^2)$ and $\bm{\beta}_*\leftarrow\bm{\beta}_*/\max(1,\norm{\bm{\beta}_*}^2)$.
%Finally, for simplicity, we sample 100 pairs of $\{(i,j)|i \in \mathcal{N}^\mathcal{I}_u \wedge  j \in \mathcal{I} \char`\\ \mathcal{N}^\mathcal{I}_u\}$ for each user $u\in\mathcal{U}$ for training.

%\textcolor{red}{\noindent\textbf{Time Complexity.}}

%\begin{table}[h]
%%	\vspace{-1ex}
%%	\small
%	%		\sffamily\raggedright\arraybackslash
%	\centering
%	\caption{Data Statistics. (Rat. denotes the range of ratings, and \#Cat. denotes the number of unique item categories.)}
%%	\vspace{-2ex}
%	\label{datastatistics}
%%	\def\arraystretch{0.88}
%	\bgroup
%	%		\begin{tabular}{p{1.95cm}|p{0.9cm}p{1.0cm}p{1.2cm}}
%	\begin{tabular}{c|NNc|J||cL}
%		\specialrule{.1em}{.1em}{.1em}
%		Dataset         & \#Users & \#Items. & \#Inter. & Density & Rat.& \#Cat.\\
%		\specialrule{.1em}{.1em}{.1em}
%		Delicious    & 1,050  & 1,196      & 7,698   &  0.61\%  & - & -\\
%		Tradesy    & 3,352  & 5,547      & 32,710   &  0.18\%  & - & -\\
%		Ciao         & 6,760  & 11,166     & 146,996 &  0.19\%  & 1-5 & 28\\			
%		Amazon C\&A    & 59,089  & 17,969     & 332,236 &  0.03\% & 1-5 & 45 \\
%		Bookcrossing & 19,571  & 39,702      & 605,178 &  0.07\% & 1-10 & -\\
%		Pinterest    & 55,187  & 9,329     & 1,462,895 &  0.28\% & - & - \\
%		Flixster    & 69,482  & 25,687     & 8,000,690 &  0.45\% & 0.5-5.0 & - \\
%		\specialrule{.1em}{.1em}{.1em}
%	\end{tabular}
%	\egroup
%%	\vspace{-3ex}
%\end{table}

%\vspace{-1ex}
\section{Experiments}
%\vspace{-1ex}
The experiments are designed to answer the following research questions (RQs):
\begin{itemize}[leftmargin=0.5cm]
	\item \textbf{RQ 1 } How does~\propose~perform compared with other state-of-the-art competitors?
	\item \textbf{RQ 2 } Are the newly introduced regularizers tailored to~\propose~beneficial for the performance of~\propose?
	\item \textbf{RQ 3 } Do the user--item specific translation vectors indeed translate the users close to their corresponding items (as illustrated in Figure~\ref{fig:explanation})?
	\item \textbf{RQ 4 } What is encoded in the translation vectors?
	\vspace{-0.5ex}
	\begin{itemize}
		\item \textit{Intensity / Heterogeneity} of user--item relationships.
	\end{itemize}
	%		\item \textbf{RQ 4} Is the \textit{strength} of user--item interactions properly encoded in the translation vectors?
	%		\item \textbf{RQ 5} Is the \textit{diversity} of user--item interactions properly encoded in the translation vectors?
%	\item \textbf{RQ 5 } How do the hyperparameters, i.e., the margin $\gamma$ and the embedding dimensions $K$, affect the model performance?
\end{itemize}

\begin{table}[h]
	%	\vspace{-1ex}
	%	\small
	%		\sffamily\raggedright\arraybackslash
	\centering
	\caption{Data Statistics. (Rat. denotes the range of ratings, and \#Cat. denotes the number of unique item categories.)}
	%		\vspace{-1ex}
	\label{datastatistics}
	\bgroup
	%		\begin{tabular}{p{1.95cm}|p{0.9cm}p{1.0cm}p{1.2cm}}
	\begin{tabular}{c|AAc|c|cc}
		\specialrule{.1em}{.1em}{.1em}
		Dataset         & \#Users & \#Items. & \#Inter. & Density & Rat.& \#Cat.\\
		\specialrule{.1em}{.1em}{.1em}
		Delicious    & 1,050  & 1,196      & 7,698 & 0.61\% & - & -\\
		Tradesy    & 3,352  & 5,547      & 32,710  & 0.13\% & - & -\\
		Ciao         & 6,760  & 11,166     & 146,996 &0.19\%  & 1-5 & 28\\			
		Amazon  & 59,089  & 17,969     & 332,236 & 0.03\% & 1-5 & 45 \\
		Bookcr& 19,571  & 39,702      & 605,178 &0.08\% & 1-10 & -\\
		Pinterest    & 55,187  & 9,329     & 1,462,895 &0.28\% & - & - \\
		Flixster    & 69,482  & 25,687     & 8,000,690 &0.45\% & 0.5-5.0 & - \\
		\specialrule{.1em}{.1em}{.1em}
	\end{tabular}
	\egroup
	\vspace{-1ex}
\end{table}

%\medskip
\noindent{\textbf{Datasets}.}
We evaluate our proposed method on \textit{seven} real-world datasets: 
Delicious
%\footnote{https://grouplens.org/datasets/hetrec-2011/}
\cite{Cantador:RecSys2011},
Tradesy
\cite{he2016vbpr},
Ciao
%\footnote{https://www.cse.msu.edu/~tangjili/datasetcode/ciao.zip}
\cite{tang-etal12a},
Amazon Cellphone and Accessories (Amazon C\&A)
%\footnote{http://jmcauley.ucsd.edu/data/amazon/}
\cite{he2016ups},
Bookcrossing
%\footnote{http://www2.informatik.uni-freiburg.de/~cziegler/BX/}
\cite{ziegler2005improving},
Pinterest
\cite{geng2015learning},
and Flixster
\cite{jamali2010matrix}.
Delicous, Tradesy and Pinterest datasets contain implicit feedback records, i.e., bookmark, purchased and pinning history.
Ciao and Amazon C\&A datasets contain rating information of users given to items, and also contain item category information, which will later be used in our experiments pertaining to verifying the heterogeneity of user--item relationships.
Bookcrossing dataset contains both the rating (from 1 to 10) information and the implicit feedback (denoted as 0) from users on items. 
%We similarly regard the observed ratings as implicit feedback together with the given implicit feedback. 
Flixster dataset contains the rating (from 0.5 to 5.0 by an interval of 0.5) information.
For datasets with rating information, we regard each observed rating as an implicit feedback record as done in previous research~\cite{rendle2009bpr,he2017neural,he2016vbpr,he2017translation,he2016ups,he2016fast,kabbur2013fism}.
We use several explicit feedback datasets and convert them into implicit feedback to be able to experimentally ascertain that the translation vectors indeed encode the intensity of user--item relationships; this is not possible with pure implicit feedback datasets as the interactions are not labeled.
We include Bookcrossing and Flixster datasets with 10-level ratings in addition to the 5-level ratings (Ciao and Amazon C\&A) to verify that~\propose~can better infer knowledge regarding the intensity of user--item relationships when finer grained user preferences are given.
We remove users and items having fewer than five ratings. The statistics of the preprocessed datasets used in our experiments are summarized in Table~\ref{datastatistics}.
%	\paragraph{Evaluation Protocol and Metrics.}

\medskip
\noindent{\textbf{Evaluation Protocol and Metrics.}}
%We adopt the widely used leave-one-out evaluation protocol, where  the last interacted item for each user is held out for test, and the rest is used for training. Since it is time consuming to rank all the items in $\mathcal{I}$, we sample 99 un-interacted items for each user and compute the ranking scores for 100 items including the test item~\cite{he2017neural,xuedeep}. For each user we additionally hold out the last interacted item from training data for validation, and tune the hyperparameters for each baseline. As we focus on recommendations for implicit feedback, we employ three ranking metrics widely used for evaluating the performance of recommender systems: \textit{Hit Ratio (HR@N), Normalized Discounted Cumulative Gain (NDCG@N),} and \textit{Mean Reciprocal Rank (MRR@N)}. \textit{HR@N} measures whether the item is presented in the top-N list, whereas \textit{NDCG@N} and \textit{MRR@N} are  position-aware ranking metrics that assign higher scores to hits at top ranks. 
We employ the widely used leave-one-out evaluation protocol~\cite{he2016vbpr,he2017neural,he2016fast,rendle2009bpr,he2017translation,xuedeep,he2016ups}, whereby  the last interacted item for each user is held out for testing, and the rest are used for training. Since it is time consuming to rank all the items in $\mathcal{I}$, we sample 99 items for each user that the user had not interacted with, and compute the ranking scores for those items plus the user's test item (100 in total  for each user) ~\cite{he2017neural,xuedeep}. For each user, we additionally hold out the last interacted item from the training data for validation set on which we tune the hyperparameters for all the methods.
As we are focused on recommendations for implicit feedback, we employ two ranking metrics widely used for evaluating the performance of recommender systems~\cite{he2017translation,he2017neural}: {hit ratio (H@N)} and {normalized discounted cumulative gain (N@N)}. H@N measures whether the item is present in the top-N list, whereas N@N is a position-aware ranking metric that assigns higher scores to hits at upper ranks. 

\begin{table*}[t]
	\centering
	\caption{Test performance of different methods. Best results are in bold face. (\textit{Imp.} denotes the improvement of~\propose~over the best competitor, which is CML.)}
	\vspace{-1ex}
	\label{tab:performance}
	\def\arraystretch{0.95}
	\begin{tabular}{Y|c|BBBBBBB||DDD|c}
		\specialrule{.1em}{.1em}{.1em}
		Datasets                      & Metrics & BPR    & FISM   & AoBPR  & eALS   & CDAE   & NeuMF    & CML    & \proposedot & \proposealt & \propose & \textit{Imp.} \\
		\specialrule{.1em}{.1em}{.1em}
		\multirow{4}{*}{\rotatebox[origin=c]{90}{Delicious}}    & H@10    & 0.1981 & 0.2203 & 0.2243 & 0.1992 & 0.1319 & 0.1164 & 0.2470  & 0.2150          & 0.2174       & \textbf{0.2586}  & 4.70\%    \\
		& H@20    & 0.3177 & 0.3391 & 0.3602 & 0.2942 & 0.2414 & 0.2171 & 0.3649 & 0.3377         & 0.3084       & \textbf{0.3786}  & 3.75\%    \\
		& N@10    & 0.1122 & 0.1124 & 0.1114 & 0.1035 & 0.0674 & 0.0558 & 0.1389 & 0.1101         & 0.1281       & \textbf{0.1475}  & 6.19\%    \\
		& N@20    & 0.1418 & 0.1424 & 0.1452 & 0.1271 & 0.0949 & 0.0789 & 0.1678 & 0.1412         & 0.1494       & \textbf{0.1781}  & 6.14\%    \\
		\specialrule{.1em}{.1em}{.1em}
		\multirow{4}{*}{\rotatebox[origin=c]{90}{Tradesy}}      & H@10    & 0.2481 & 0.2676 & 0.2597 & 0.2058 & 0.1652 & 0.1167 & 0.3031 & 0.2846         & 0.2648       & \textbf{0.3198}  & 5.51\%    \\
		& H@20    & 0.4174 & 0.4109 & 0.4256 & 0.3314 & 0.2867 & 0.2290  & 0.4413 & 0.4266         & 0.3823       & \textbf{0.4505}  & 2.08\%    \\
		& N@10    & 0.1248 & 0.1309 & 0.1300   & 0.1042 & 0.0831 & 0.0538 & 0.1685 & 0.1449         & 0.1466       & \textbf{0.1767}  & 4.87\%    \\
		& N@20    & 0.1673 & 0.1670  & 0.1715 & 0.1356 & 0.1136 & 0.0817 & 0.2031 & 0.1806         & 0.1760        & \textbf{0.2095}  & 3.15\%    \\
		\specialrule{.1em}{.1em}{.1em}
		\multirow{4}{*}{\rotatebox[origin=c]{90}{Ciao}}         & H@10    & 0.1569 & 0.2100   & 0.1873 & 0.1419 & 0.1770  & 0.1535 & 0.2085 & 0.2011         & 0.1991       & \textbf{0.2292}  & 9.93\%    \\
		& H@20    & 0.2811 & 0.3482 & 0.3146 & 0.2570  & 0.3153 & 0.2788 & 0.3337 & 0.3185         & 0.3270        & \textbf{0.3740}   & 12.08\%   \\
		& N@10    & 0.0751 & 0.1027 & 0.0891 & 0.0670  & 0.0862 & 0.0741 & 0.1053 & 0.1017         & 0.0989       & \textbf{0.1167}  & 10.83\%   \\
		& N@20    & 0.1063 & 0.1374 & 0.1209 & 0.0957 & 0.1208 & 0.1040  & 0.1358 & 0.1311         & 0.1309       & \textbf{0.1525}  & 12.30\%   \\
		\specialrule{.1em}{.1em}{.1em}
		\multirow{4}{*}{\rotatebox[origin=c]{90}{\begin{tabular}[x]{@{}c@{}}Book-\\crossing\end{tabular}}} & H@10    & 0.2425 & 0.2178      & 0.2563 & 0.1655 & 0.2244 & 0.2286 & 0.2885 & 0.2802         & 0.2828       & \textbf{0.3329}   & 15.39\%   \\
		& H@20    & 0.3761 & 0.3938      & 0.3916 & 0.2864 & 0.3610  & 0.3747 & 0.4053 & 0.3932         & 0.4069       & \textbf{0.4744}  & 17.05\%   \\
		& N@10    & 0.1250  & 0.1002      & 0.1338 & 0.0791 & 0.1164 & 0.1158 & 0.1663 & 0.1618         & 0.1578       & \textbf{0.1865}  & 12.15\%    \\
		& N@20    & 0.1585 & 0.1444      & 0.1676 & 0.1093 & 0.1506 & 0.1482 & 0.1956 & 0.1903         & 0.1890        & \textbf{0.2221}  & 13.55\%    \\
		\specialrule{.1em}{.1em}{.1em}
		\multirow{4}{*}{\rotatebox[origin=c]{90}{\begin{tabular}[x]{@{}c@{}}Amazon\\C\&A\end{tabular}}}       & H@10    & 0.2489 & 0.2470  & 0.2646 & 0.2161      & 0.2817 & 0.1317 & 0.3011 & 0.3003         & 0.3184       & \textbf{0.3436}  & 14.11\%   \\
		& H@20    & 0.3821 & 0.3782 & 0.3946 & 0.3480      & 0.4117 & 0.2390  & 0.4123 & 0.4184         & 0.4509       & \textbf{0.4658}  & 12.98\%   \\
		& N@10    & 0.1276 & 0.1247 & 0.1391 & 0.1064      & 0.1613 & 0.0613 & 0.1752 & 0.1648         & 0.1766       & \textbf{0.2019}  & 15.24\%   \\
		& N@20    & 0.1610  & 0.1577 & 0.1718 & 0.0739      & 0.1939 & 0.0880  & 0.2031 & 0.1945         & 0.2094       & \textbf{0.2323}  & 14.38\%   \\
		\specialrule{.1em}{.1em}{.1em}
		\multirow{4}{*}{\rotatebox[origin=c]{90}{Pinterest}}    
		& H@10    & 0.4759 & 0.4444 & 0.4921 & 0.3301 & 0.5244 & 0.4546 & 0.5378 & 0.5485         & 0.4899       & \textbf{0.5504}  & 2.34\%    \\
		& H@20    & 0.7564 & 0.6720  & 0.7618 & 0.5621 & 0.7393 & 0.6852 & 0.7771 & 0.7822         & 0.7514       & \textbf{0.8108}  & 4.34\%    \\
		& N@10    & 0.2034 & 0.2048 & 0.2143 & 0.1373 & 0.2644 & 0.2165 & 0.2558 & 0.2549         & 0.2259       & \textbf{0.2580}   & 0.86\%    \\
		& N@20    & 0.2744 & 0.2626 & 0.2827 & 0.1959 & 0.3188 & 0.2748 & 0.3166 & 0.3143         & 0.2922       & \textbf{0.3242 } & 2.40\%    \\
		\specialrule{.1em}{.1em}{.1em}
		\multirow{4}{*}{\rotatebox[origin=c]{90}{Flixster}} 
		&H@10    & 0.6836      & 0.5985      & 0.6904      & 0.6320      & 0.6797      & 0.6596      & 0.7009      & 0.6014   & 0.7123            & \textbf{0.7309}       & 4.28\%       \\
		&H@20 &0.8087&0.7597&0.8124&0.7359&0.7973&0.7816&0.8081&0.7147&0.8163& \textbf{0.8374}& 3.63\%       \\
		&N@10    & 0.3701      & 0.2794      & 0.3830      & 0.3513      & 0.4526      & 0.4588      & 0.4608      & 0.3417   & 0.4704            & \textbf{0.4986}       & 8.20\%       \\
		&N@20    & 0.4020      & 0.3206      & 0.4140      & 0.3778      & 0.4824      & 0.4895      & 0.4881      & 0.3705   & 0.4969            & \textbf{0.5257}       & 7.70\%      \\
		\specialrule{.1em}{.1em}{.1em}
	\end{tabular}
\vspace{-1ex}
\end{table*}

\medskip
\noindent{\textbf{Methods Compared. }}
As~\propose~is a pair-wise \textit{1) learning-to-rank} method based on \textit{2) metric learning} that employs \textit{3) neighborhood information}, we choose the following baselines.
\begin{enumerate}[leftmargin=0.5cm]
	\item Learning-to-rank baselines.
%	\item \textbf{WMF}~\cite{hu2008collaborative}: A pointwise weighted matrix factorization method for implicit feedback.
	\begin{itemize}
		\item Point-wise methods.
		\begin{itemize}
			\item \textbf{eALS}~\cite{he2016fast}: The state-of-the-art MF--based method for implicit feedback that non-uniformly weights the unobserved interactions based on the item popularity.
			\item \textbf{NeuMF}~\cite{he2017neural}: A pointwise neural collaborative filtering framework for implicit feedback that combines MF and multi-layer perceptron (MLP). We report the best results from among MF, MLP, and NeuMF.
		\end{itemize}
		\item Pairwise methods.
		\begin{itemize}
			\item \textbf{BPR}~\cite{rendle2009bpr}: A pairwise learning-to-rank method for implicit feedback in which observed items are assumed to be highly preferred by users to unobserved items.
			\item \textbf{AoBPR}~\cite{rendle2014improving}: An extension of BPR that samples popular items as negative feedback with a higher probability.
		\end{itemize}
	\end{itemize}
	
	\item Neighborhood--based baselines.
	\begin{itemize}
		\item \textbf{FISM}~\cite{kabbur2013fism}: A neighborhood--based recommendation method based on MF in which a user is represented by the items that the user has interacted with as in Eqn.~\ref{eqn:usernbr}.
		\item \textbf{CDAE}~\cite{wu2016collaborative}: The state-of-the-art neighborhood--based method in which the user--item relationship is computed between a user and his neighbors, i.e., the items that the user has previously interacted with.
	\end{itemize}
	\item Metric learning--based baselines \& Ablations of~\propose.
	\begin{itemize}
		\item \textbf{CML}~\cite{hsieh2017collaborative}: The state-of-the-art metric learning--based recommendation method for implicit feedback in which Euclidean distance is used for the scoring function. i.e., $s(u,i) = -\norm{\bm{\alpha}_u - {\bm{\beta}_i}}_2^2$.
%		CML is the most closely related to our proposed method.
		%	\item \textbf{\propose} Our proposed method.
		\item \textbf{\proposedot}: An ablation of~\propose~in which instead of Euclidean distance, inner product is used for the scoring function. i.e., $s(u,i) = (\bm{\alpha}_u + \bm{r}_{ui})^T{\bm{\beta}_i}$.
		\item \textbf{\proposealt}: Another ablation of~\propose~in which the translation vector is computed by $\bm{r}_{ui} = f(\bm{\alpha}_u,  \bm{\beta}_i)$. i.e., we employ the current user and item embeddings for constructing $\bm{r}_{ui}$ \textit{instead of their neighborhoods}. 
	\end{itemize}
\end{enumerate}

%\vspace{-1ex}
\medskip
\noindent{\textbf{Implementation Details}.}
%	\paragraph{Hyperparameters.}
For each data, the hyperparameters are tuned on the validation set by grid searches with $K\in\{8,16,32,64,128\}$,  $\eta\in\{0.0005,0.001,0.005,0.01,0.05,0.1\}$, $\gamma\in\{0.0,0.1,0.5,1.0$\\$,1.5,2.0,$$2.5,3.0\}$, $\lambda,\lambda_{\mathrm{nbr}},\lambda_{\mathrm{dist}}\in\{0.0,0.001,0.01,0.1\}$, where $\lambda$ is the regularization coefficient for the baseline methods. 
%All the reported results are the test performance measured using the best performing hyperparameters on the validation set.
We report the test performance measured using the hyperparameter values that give the best HR@10 on the validation set.
For reliability, we repeat our evaluations five times with different random seeds for the model initialization, and we report mean test errors.
We fix the number of samples in a mini-batch to 1000 for mini-batch SGD.
%and use Adam~\cite{kingma2014adam} to control the learning rate $\eta$. 
%For each method, we choose the hyperparameter values that give the best {HR@10} value on the validation dataset. 

%\vspace{-1.5ex}
\subsection{Performance Analysis}
\label{exp:quant}
\subsubsection{\textbf{Recommendation performance (RQ 1)}}
Table~\ref{tab:performance} shows the performance of different methods in terms of various ranking metrics.
%averaged over five runs that used different random seeds for the model initializations. 
We have the following observations from Table~\ref{tab:performance}. 1) We observe that CML outperforms the MF--based competitors (BPR, FISM, AoBPR, eALS, and NeuMF). This is consistent with the previous work~\cite{hsieh2017collaborative}, indicating that metric learning approaches overcome the inherent limitation of MF by learning a metric space wherein the triangle inequality is satisfied.
2) We observe that~\propose~considerably outperforms the state-of-the-art competitor, namely CML, by up to 17.05\% (achieved for HR@20 on Bookcrossing dataset). 
This verifies the benefit of the translation vectors that translate each user toward items according to the user's relationships with those items. 
%Moreover,~\proposealt~also generally outperforms CML. This verifies the benefit of the translation vectors that puts each user close to the target items. 
3)~\proposealt~generally performs worse than CML, which implies that the translation vectors should be carefully designed, or else the performance will rather deteriorate.
4) The superior performance of~\propose~over~\proposealt~confirms that incorporating the neighborhood information is indeed crucial in collaborative filtering~\cite{koren2008factorization,kabbur2013fism,wu2016collaborative}. 
5)~\proposedot~generally performs worse than CML, which verifies that the inner product operation limits the performance despite the benefit of the translation mechanism adopted to~\proposedot.
%\textcolor{red}{TODO:~\proposedot~vs~rest}
6) By further comparing HR@10 of~\proposealt~on Delicious and Bookcrossing datasets (0.2174 and 0.2828 in Table~\ref{tab:performance}) with HR@10 of~\propose~without any regularization (0.2388 and 0.2983 in Table~\ref{tab:reg} when $\lambda_{\mathrm{dist}}=\lambda_{\mathrm{nbr}}=0.0$), we observe that~\propose~without the regularizers still outperforms~\proposealt. This again verifies that the neighborhood information itself is indeed beneficial even without the help of the neighborhood regularizer. 
%Note that we tried swapping the user and item
%6) It is noteworthy that the performance improvements on Delicious dataset are smaller than those on the rest of the datasets. We attribute this to the fact that Delicious dataset is a small dataset containing relatively fewer items than the other datasets. The user--item interactions pertaining to CF are less complex in such a dataset, and thus projecting a user to a single point, as is done in CML, yields comparable results.

%\vspace{-1ex}
\begin{table}[t]
	\centering
%	\small
	\caption{Effect of the regularization coefficients.}
%	\vspace{-1ex}
	\label{tab:reg}
	\def\arraystretch{0.95}
	\begin{tabular}{cc|cccc|}	
		\hline
		\multicolumn{2}{|c|}{\multirow{2}{*}{\begin{tabular}[x]{@{}c@{}}Delicious\\(HR@10)\end{tabular}}}                    & \multicolumn{4}{c|}{$\lambda_{\mathrm{nbr}}$}                                                                                \\ \cline{3-6} 
		\multicolumn{2}{|c|}{}                                     & \multicolumn{1}{c}{0.0} & \multicolumn{1}{c}{0.001} & \multicolumn{1}{c}{0.01} & \multicolumn{1}{c|}{0.1} \\ \cline{1-6}
		\multicolumn{1}{|c|}{\multirow{4}{*}{\rotatebox[origin=c]{90}{$\lambda_{\mathrm{dist}}$}}} 
		& 0.0   & 0.2388            & 0.2388                     & 0.2401                    & 0.2454                   \\
		\multicolumn{1}{|c|}{}                           & 0.001 & 0.2375            & 0.2401                     & 0.2401                    & 0.2454                   \\ 
		\multicolumn{1}{|c|}{}                           & 0.01  & 0.2401            & 0.2401                     & 0.2427                    & 0.2467                   \\ 
		\multicolumn{1}{|c|}{}                           & 0.1   & 0.2520            & 0.2520         			  & 0.2533                    & \textbf{0.2586}                   \\ 
		\hline
		\hline

		\multicolumn{2}{|c|}{\multirow{2}{*}{\begin{tabular}[x]{@{}c@{}}Bookcrossing\\(HR@10)\end{tabular}}}                    & \multicolumn{4}{c|}{$\lambda_{\mathrm{nbr}}$}                                                                                \\ \cline{3-6} 
		\multicolumn{2}{|c|}{}                                     & \multicolumn{1}{c}{0.0} & \multicolumn{1}{c}{0.001} & \multicolumn{1}{c}{0.01} & \multicolumn{1}{c|}{0.1} \\ \cline{1-6}
		\multicolumn{1}{|c|}{\multirow{4}{*}{\rotatebox[origin=c]{90}{$\lambda_{\mathrm{dist}}$}}} 
		& 0.0   							& 0.2983            & 0.3033                     & 0.3227                    & 0.3185                   \\
		\multicolumn{1}{|c|}{}      & 0.001 & 0.3038            & 0.3074                     & 0.3236                    & 0.3181                  \\ 
		\multicolumn{1}{|c|}{}      & 0.01  & 0.3213            & 0.3219                     & 0.3264                    & 0.3187                   \\ 
		\multicolumn{1}{|c|}{}      & 0.1   & \textbf{0.3329}           & 0.3329         			 & 0.3324                    & 0.3151                   \\ 
		\hline
	\end{tabular}
	\vspace{-2ex}
\end{table}

%\vspace{-1ex}
\smallskip
\subsubsection{\textbf{Benefit of regularizers (RQ 2)}}
\label{sec:exp_reg}
Table~\ref{tab:reg} shows the effect of the regularization coefficients on the performance of~\propose~on Delicious and Bookcrossing datasets, where $\lambda_{\mathrm{nbr}}$ and $\lambda_{\mathrm{dist}}$ denote the strengths of the neighborhood regularizer ($reg_{\mathrm{nbr}}$) and of the distance regularizer ($reg_{\mathrm{dist}}$), respectively. Larger values imply a stronger contribution of the regularizer to the model, and $\lambda_*=0.0$ indicates no regularization.
We have the following observations: 1)
Both regularizers are indeed beneficial for the model performance, and their impact varies across different datasets. %i.e., the neighborhood regularizer is helpful on Delicious dataset, whereas it deteriorates the performance on Ciao dataset. 
2) Although $reg_{\mathrm{nbr}}$ is beneficial, its impact on the model performance decreases as the $reg_{\mathrm{dist}}$ dominates the model; i.e., as $\lambda_{\mathrm{dist}}$ increases.
3) $reg_{\mathrm{dist}}$  is more helpful for the model performance compared with $reg_{\mathrm{nbr}}$; the benefit of $reg_{\mathrm{dist}}$ becomes more clear on Bookcrossing dataset in which the user--item relations are more complex compared with Delicious dataset. This confirms that explicitly pulling each translated user toward all of the positive items rather than merely pushing negative items away from each user helps to model the complex user--item interactions, which aligns with our motivation for the distance regularizer $reg_{\mathrm{dist}}$ as mentioned in Section~\ref{subsec:reg_dist}.

%\begin{table}[h]
%	\centering
%	%	\small
%	\caption{Analysis on the user--item specific translation vectors. (\textit{Observed/Unobserved} denotes results on observed/unobserved user--item interactions.)}
%		\vspace{-1ex}
%	\label{tab:posneg}
%	\def\arraystretch{0.95}
%	\begin{tabular}{c|cc|c}
%		\hline
%		Dataset & \textit{Observed} & \textit{Unobserved} & Difference \\
%		\hline
%		\hline
%		Delicious    & 64.63\%  & 43.75\%  & 20.88\%    \\
%		Tradesy      & 56.02\%  & 43.01\%  & 13.01\%    \\
%		Ciao         & 54.63\%  & 38.42\%  & 16.21\%    \\
%		Bookcrossing & 55.42\%  & 35.57\%  & 19.85\%    \\
%		Amazon C\&A    & 75.57\%  & 31.96\%  & 43.62\%   \\		
%		Pinterest    & 36.25\%  & 33.08\%  & 3.17\%    \\
%		Flixster    & 22.24\%  & 2.88\%  & 19.36\%    \\
%		\hline
%	\end{tabular}
%		\vspace{-3ex}
%\end{table}

%\begin{table}[h]
%	\centering
%	%	\small
%	\caption{Analysis on the user--item specific translation vectors. (\textit{Observed/Unobserved} denotes results on observed/unobserved user--item interactions.)}
%	\vspace{-1ex}
%	\label{tab:posneg}
%	\def\arraystretch{0.95}
%	\begin{tabular}{c|cc}
%		\hline
%		Dataset & \textit{Observed} & \textit{Unobserved} \\
%		\hline
%		\hline
%		Delicious    & 64.63\%  & 43.75\%     \\
%		Tradesy      & 56.02\%  & 43.01\%     \\
%		Ciao         & 54.63\%  & 38.42\%      \\
%		Bookcrossing & 55.42\%  & 35.57\%    \\
%		Amazon C\&A    & 75.57\%  & 31.96\%   \\		
%		Pinterest    & 36.25\%  & 33.08\%    \\
%		Flixster    & 22.24\%  & 2.88\%     \\
%		\hline
%	\end{tabular}
%	\vspace{-3ex}
%\end{table}

\begin{table}[h]
	\centering
	%	\small
	\caption{Analysis on the user--item specific translation vectors. (\textit{Obs./Unobs.} denotes results on observed/unobserved user--item interactions.)}
	\vspace{-1ex}
	\label{tab:posneg}
	\begin{tabular}{c|cc||c|cc}
		\hline
		Dataset & \textit{Obs.} & \textit{Unobs.}&Dataset & \textit{Obs.} & \textit{Unobs.} \\
		\hline
		\hline
		Delicious    & 64.63\%  & 43.75\% &Amazon   & 75.57\%  & 31.96\%    \\
		Tradesy      & 56.02\%  & 43.01\% &Pinterest    & 36.25\%  & 33.08\%    \\
		Ciao         & 54.63\%  & 38.42\% &Flixster    & 22.24\%  & 2.88\%         \\
		Bookcr. & 55.42\%  & 35.57\%   &&& \\
		\hline
	\end{tabular}
	\vspace{-2ex}
\end{table}

\subsection{Qualitative Evaluations}
\vspace{-0.5ex}
\label{exp:qual}
\subsubsection{\textbf{Benefit of translation vectors (RQ 3)}} 
%Recall that a limitation of CML is that each user is projected to a single point in the metric space, which fails to model the diversity and the strength of user--item interactions.
In this section, we conduct experiments to verify whether the translation vectors learned by~\propose~indeed translate each user closer to the observed (positive) items as in \textbf{Toy example} illustrated in Figure~\ref{fig:explanation}.
To this end, for each user $u$ and his observed item $i$, we check whether the following holds: 
\begin{equation}
	\small
	\label{eqn:cond}
	\norm{\bm{\alpha}_u-\bm{\beta}_i}^2_2 > \norm{\bm{\alpha}_u+\bm{r}_{ui}-\bm{\beta}_i}^2_2
\end{equation}
That is, we expect the distance between the item embedding vector $\bm{\beta}_i$ of observed item $i$ and the translated user embedding vector $(\bm{\alpha}_u + \bm{r}_{ui})$ to be smaller than the distance between the item embedding vector $\bm{\beta}_i$ and the user embedding vector $\bm{\alpha}_u$ before translation.
We calculate the percentage of observed/unobserved
%\textit{Positive (Negative)} denotes the portion of 
user--item interaction pairs that satisfy Equation~\ref{eqn:cond} among all possible observed/unobserved pairs. 
We sample as many unobserved items as the number of observed items for each user for the comparisons. 
Here, we expect more observed pairs to satisfy Equation~\ref{eqn:cond} than unobserved pairs.
Table~\ref{tab:posneg} shows the results on the seven datasets. 

We observe that users are generally translated closer toward their observed (positive) items, and at the same time translated farther away from the unobserved (negative) items. For instance, for Amazon C\&A dataset, 75.57\% of the observed user--item interactions satisfy Equation~\ref{eqn:cond}, whereas only 31.96\% of the unobserved interactions satisfy it, which conversely implies that users in 68\% ($\approx 100-31.96$) of the unobserved interactions translate users farther away from the unobserved items. We also note that for Flixster dataset, although only 22.24\% of the observed interactions satisfy Equation~\ref{eqn:cond}, the overwhelming majority of the unobserved interactions, i.e., 97\% ($\approx 100-2.88$), violate it, which means that the effect of users being translated away from the unobserved items results in placing the translated users relatively closer toward their observed items.
%dominates the model without compromising the overall recommendation performance.
Hence, we argue that by simultaneously translating a user toward his relevant items and away from his irrelevant items,~\propose~achieves superior recommendation performance.

%We observe that~\propose~is properly trained so that each translated user is placed closer to the observed (positive) items than to the unobserved (negative) items. Notably, the difference between the observed and unobserved cases is the largest on Amazon C\&A dataset and the smallest on Pinterest dataset, implying that~\propose~is best trained on the Amazon C\&A whereas worst trained on Pinterest dataset. These results are supported by Table~\ref{tab:performance} in which the performance improvements of~\propose~is large on the Amazon C\&A dataset and the smallest on Pinterest dataset.

%We note that the as these numbers are relative measurements, 
%We note that the purpose of this experiment is to show that there is a general trend of translating a user closer toward his observed items in comparison to his unobserved items, not to show that a user is always translated closer toward his every observed items; some users may be translated farther away from his observed items, but still according to the relative intensity of his relationship among the observed items.

\medskip
\subsubsection{\textbf{What is encoded in the translation vectors? (RQ 4)}}
\label{sec:int_het}
In this section, we investigate why~\propose~outperforms the state-of-the-art methods. For this purpose, we conduct various experiments with the translation vectors to verify our claims that the translation vectors encode 
the \romannum{1}) {\textbf{intensity}}, and the \romannum{2}) {\textbf{heterogeneity}} of user--item relationships in implicit feedback.
%the \textbf{strength} and the \textbf{diversity} of the user--item interactions. 

\begin{table}[t]
	\centering
	%	\small
	\caption{Rating classification accuracy using translation vectors. (\textit{Rand} denotes a random classifier, and RF denotes the random forest classifier~\cite{breiman2001random}.)}
	%	\vspace{-2ex}
	\label{tab:classification_rat}
	%	\def\arraystretch{0.95}
	%       \begin{tabular}{|I|T|H|T|H|T|H|T|H|}
	\begin{tabular}{|I|L|S|L|S|L|S|L|S|}
		\hline
		\multirow{2}{*}{
			\begin{tabular}[x]{@{}c@{}}Acc.(\%)\end{tabular}}		&\multicolumn{2}{c|}{Ciao}	&\multicolumn{2}{c|}{Amazon} &\multicolumn{2}{c|}{BookCr.\footnotemark} &\multicolumn{2}{c|}{Flixster} \\      
		\cline{2-9}
		& \textit{Rand}   										 & RF                   & \textit{Rand}  										 & RF					& \textit{Rand}   										& RF        & \textit{Rand}   										& RF     		\\                      
		\hline
		\hline
		CML		     &\tikz[overlay] \draw (0,0.7em)--(0,-0.1em);     & 50.3   &\tikz[overlay] \draw (0,0.7em)--(0,-0.1em);     & 50.1 	&\tikz[overlay] \draw (0,0.7em)--(0,-0.1em);    & 39.1 &\tikz[overlay] \draw (0,0.7em)--(0,-0.1em);    & 20.5       		\\
		\proposeemb  &19.9                                         & 50.3            	&20.1                               & 50.3    &13.8                               			& 40.1 &10.0                               			& 20.5        		\\
		\propose     &\tikz[overlay] \draw (0,0.7em)--(0,-0.1em);     &\textbf{53.0}    	&\tikz[overlay] \draw (0,0.7em)--(0,-0.1em);     & \textbf{50.8} 	&\tikz[overlay] \draw (0,0.7em)--(0,-0.1em);    &\textbf{43.7}&\tikz[overlay] \draw (0,0.7em)--(0,-0.1em);    &\textbf{23.4}		\\
		\hline
		\hline
		\textit{vs. CML} 		&- 		 & 5.3 				&- 								 & 1.5				& -			& 11.7 & -			& 14.2				\\
		\hline
	\end{tabular}
	\vspace{-2ex}
\end{table}
\footnotetext[3]{We regard ratings of less than 4 as a single class as they account for only 3.86\%.}

\smallskip
\noindent\textbf{\textit{Intensity} of user--item interactions.}
With regard to the first claim, i.e., intensity, we assume that the rating information is a proxy for the intensity of user--item relationships. In other words, the higher the rating of an item given by a user, the higher the intensity of the user--item relationship. Since Ciao, Amazon C\&A, Bookcrossing and Flixster datasets contain rating information, we use them to study whether we can conversely infer some knowledge about ratings using the translation vectors \textit{even though the ratings are not utilized during the model training}. 
To this end, we perform rating classification to predict the rating of user $u$ on item $i$ given the user--item specific translation vector $\bm{r}_{ui}$ as input; for instance, there are five classes for datasets with ratings ranging from 1 to 5. Note that we perform classification instead of regression to clearly emphasize the difference between the numbers; the RMSE values of regression usually differ in the second decimal place~\cite{koren2008factorization}.
%Note that the number of classes is equivalent to different number of rating classes.
%Each rating is considered as a class as ratings are integers. 
To ascertain the benefit of the translation vectors learned by~\propose, i.e., $\bm{r}^{\propose}_{ui}$, we compare the rating classification performance with that of CML. Since CML does not model the translation vectors~\cite{hsieh2017collaborative}, we alternatively define the translation vectors for CML, i.e., $\bm{r}^{\mathrm{CML}}_{ui}\coloneqq (\bm{\alpha}_u-\bm{\beta}_i)$, where $\bm{\alpha}_u$ and $\bm{\beta}_i$ are both trained by CML. Likewise, we additionally define synthetic translation vectors $\bm{r}_{ui}^{\propose^{\mathrm{emb}}}\coloneqq (\bm{\alpha}_u-\bm{\beta}_i)$ for~\propose~, where $\bm{\alpha}_u$ and $\bm{\beta}_i$ are both trained by~\propose, and name this method~\proposeemb. Table~\ref{tab:classification_rat} summarizes the classification results\footnote{\label{note1}We balance the class distribution by randomly sampling a fixed number of samples (num. samples in the minority class) from each class. We then perform 5-fold cross validation. We use the default setting of the random forest classifier in Scikit-learn~\cite{scikit-learn}, but the accuracies could be further improved by tuning the classifier.}.
%\footnote{\label{note1}To cope with the class imbalance issue, we sample from each class as many samples as the number of ratings in the smallest class. We perform 5-fold cross validation.}.
We observe that the rating classification accuracy on translation vectors of~\propose~outperforms that of CML and~\proposeemb, and that the improvement is the greatest on Flixster dataset. 
This implies that the \textbf{intensity} of user--item relationships is encoded in the translation vectors learned by~\propose, which explains the superior performance of~\propose~as shown in Table~\ref{tab:performance}.
%However, we note that the actual classification accuracies are not sufficiently high,
%reaching only 53\% in Ciao dataset, motivating us to further investigate the translation vectors with regard to the intensity of user--item relationships.
%, which is in line with the previous results in Table~\ref{tab:posneg:ratingwise}. 
However, we note that the improvements of classification accuracies compared with CML are not sufficiently large on Ciao and Amazon C\&A datasets, reaching only 5.3\% and 1.5\%, respectively.
This motivates us to further investigate the translation vectors with regard to the intensity of user--item relationships.

\begin{table}[t]
	\centering
%	\small
	\caption{Analysis on the user--item specific translation vectors regarding the ratings. (\textit{Eqn~\ref{eqn:cond}}. denotes the percentage of interactions that satisfy Equation~\ref{eqn:cond}, and Ptn. denotes the portion of each class.)}
%	\vspace{-1ex}
	\label{tab:posneg:ratingwise}
	\begin{tabular}{|J||cNNNNNNN}
		\cline{1-6}
		& \multicolumn{5}{c|}{Rating}                                                                                                                                                             &                        \\
		\cline{1-6}
		Ciao         & \multicolumn{1}{c}{1}            & \multicolumn{1}{c}{2} & \multicolumn{1}{c}{3}   & \multicolumn{1}{c}{4} & \multicolumn{1}{c|}{5}   &  & \multicolumn{2}{l}{\multirow{6}{*}{}}           \\
		\cline{1-6}
		Eqn~\ref{eqn:cond}         & 61.5\%                          & 51.4\%               & 55.4\%                 & 52.2\%               & \multicolumn{1}{c|} {55.4\%}                 &                        & \multicolumn{2}{l}{}                            \\
		Ptn.          & 4.8\%                           & 5.1\%                & 11.4\%                 & 29.0\%               & \multicolumn{1}{c|}{49.7\%}                 &                & \multicolumn{2}{l}{}                            \\
%		\cline{1-6}
%		\cline{1-6}
		\hhline{|======|}
		Amazon       & \multicolumn{1}{c}{1}            & \multicolumn{1}{c}{2} & \multicolumn{1}{c}{3}   & \multicolumn{1}{c}{4} & \multicolumn{1}{c|}{5}   &  & \multicolumn{2}{l}{}                            \\
		\cline{1-6}
		Eqn~\ref{eqn:cond}         & 76.7\%                          & 76.3\%               & 75.7\%                 & 75.2\%               & \multicolumn{1}{c|} {75.4\%}                 &                        & \multicolumn{2}{l}{}                            \\
		Ptn.          & 7.0\%                           & 5.7\%                & 10.7\%                   & 20.1\%               & \multicolumn{1}{c|} {56.5\%}                 &                & \multicolumn{2}{l}{}                            \\
%		\cline{1-5}
%		\cline{1-8}
%		\hhline{======}
		\hhline{|======|}
		\hhline{~~~~~~--}
		BookCr. & \multicolumn{1}{c}{1-4}     & \multicolumn{1}{c}{5} & \multicolumn{1}{c}{6}   & \multicolumn{1}{c}{7} & \multicolumn{1}{c}{8}   & \multicolumn{1}{c}{9}  & \multicolumn{1}{c|}{10} &  \\
		\cline{1-8}
		Eqn~\ref{eqn:cond}         & 55.3\%                           & 52.7\%                & 55.2\%                  & 56.1\%                & 57.2\%                  & 58.4\%                 & \multicolumn{1}{c|} {58.8\%}                &                        \\
		Ptn.          & 3.8\%                            & 10.3\%                & 7.9\%                   & 17.0\%                & 24.5\%                  & 17.3\%                 & \multicolumn{1}{c|} {19.2\%}                 & \\
%		\cline{1-8}
%		\cline{1-7}
		\cline{7-8}
		\hhline{|=======|}
		Flixster     & \multicolumn{1}{c}{0.5-2.5} & \multicolumn{1}{c}{3.0} & \multicolumn{1}{c}{3.5} & \multicolumn{1}{c}{4.0} & \multicolumn{1}{c}{4.5} & \multicolumn{1}{c|}{5.0}  &  & \multirow{3}{*}{}      \\
		\cline{1-7}
		Eqn~\ref{eqn:cond}         & 19.6\%                           & 19.9\%                & 19.9\%                  & 22.2\%                & 25.7\%                  & \multicolumn{1}{c|} {27.2\%}                 &                        &                        \\
Ptn.          			   & 17.3\%                           & 17.0\%                & 16.8\%                  & 19.6\%                & 10.1\%                  & \multicolumn{1}{c|} {19.2\%}                 &                 &                       \\
		\cline{1-7}
	\end{tabular}
\vspace{-1ex}
\end{table}

%To this end, 
To further study the translation vectors,
we group the results of the observed user--item interactions from Table~\ref{tab:posneg} according to their ground truth ratings, and calculate the percentage of interactions that satisfy Equation~\ref{eqn:cond} in each rating group.
%accuracy in terms of Equation~\ref{eqn:cond} on each rating group. 
Table~\ref{tab:posneg:ratingwise} shows the results. 
Under the assumption that higher ratings imply a higher intensity of user--item relationships, we expect more observed interactions to satisfy Equation~\ref{eqn:cond} in higher rating groups.
%the higher the ratings.the more interactions satisfy Equation~\ref{eqn:cond}, 
%That is to say, it is more likely for a translated user to be put closer to his highly relevant items than to less relevant items.
%We have the following observations:
%1) The results on Ciao and Amazon C\&A datasets do not agree with our above expectation, 
However, we observe that the results on Ciao and Amazon C\&A datasets do not agree with our expectation, which explains the relatively small improvements in Table~\ref{tab:classification_rat}. We conjecture that it is inherently challenging to infer users' fine-grained preferences from the user--item interactions of Ciao and Amazon C\&A datasets, because 1) the range of the ratings is relatively small (integers from 1 to 5), and more importantly, 2) the majority (over 75\%) of interactions belong to ratings from 4 to 5.
In contrast, for Bookcrossing\footnote{Note that for Bookcrossing dataset, the number of observed interactions whose ratings are less than 4 account for a small proportion (3.86\%), and are therefore negligible.} and Flixster datasets whose ratings are in a wider range (from 1 to 10, and from 0.5 to 5.0, respectively), and relatively evenly distributed throughout the rating classes, the percentage of interactions satisfying Equation~\ref{eqn:cond} increases as the ratings increase from 5 to 10 (for Flisxter, from 3.0 to 5.0). These results on Bookcrossing and Flixster datasets (Table~\ref{tab:posneg:ratingwise}) are in line with the results in Table~\ref{tab:classification_rat}, where the relative improvement of the classification results of~\propose~is the highest for Bookcrossing and Flixster datasets.
To summarize our findings from Tables~\ref{tab:classification_rat} and~\ref{tab:posneg:ratingwise}, 
%we conclude that the intensity of user--item interactions is indeed encoded in the translation vectors, while it becomes more clear for datasets from which users' preferences can be easily inferred. e.g., Bookcrossing.
we conclude that the intensity of user--item relationships is indeed encoded in the translation vectors, and that it becomes clearer with datasets from which users' preferences can be more precisely inferred, e.g., Bookcrossing and Flisxter datasets.

However, up to this point, it is still uncertain why the improvements of~\propose~on Ciao and Amazon C\&A datasets are considerable in terms of the quantitative evaluation (Table~\ref{tab:performance}), even though the rating information (intensity) is not as clearly encoded in the translation vectors as in the case of Bookcrossing and Flixster dataset. 
Hence, in the following section, we study whether the translation vectors of~\propose~trained on Ciao and Amazon C\&A datasets encode meaningful information other than the rating information.

\begin{figure*}[t]
	\centering
	\begin{minipage}{0.6\textwidth}
		%		\vspace{-2ex}
		\centering\captionsetup{width=0.95\linewidth}
		\begin{subfigure}{0.495\textwidth}
			\centering
			\includegraphics[width=1\linewidth]{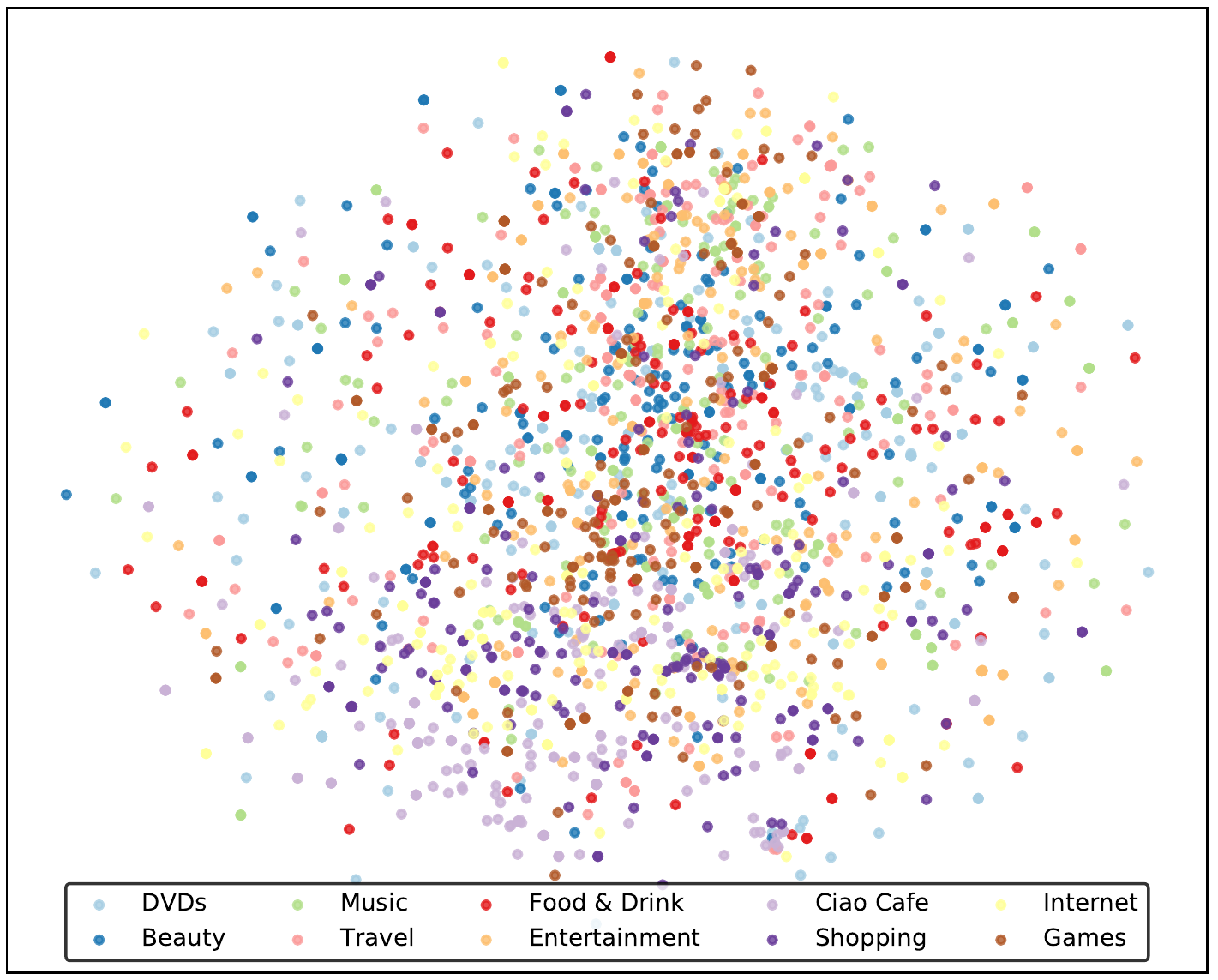}
			\caption{Visualization of $\bm{r}^{CML}_{ui}$}
		\end{subfigure}
		\begin{subfigure}{0.495\textwidth}
			\centering
			\includegraphics[width=1\linewidth]{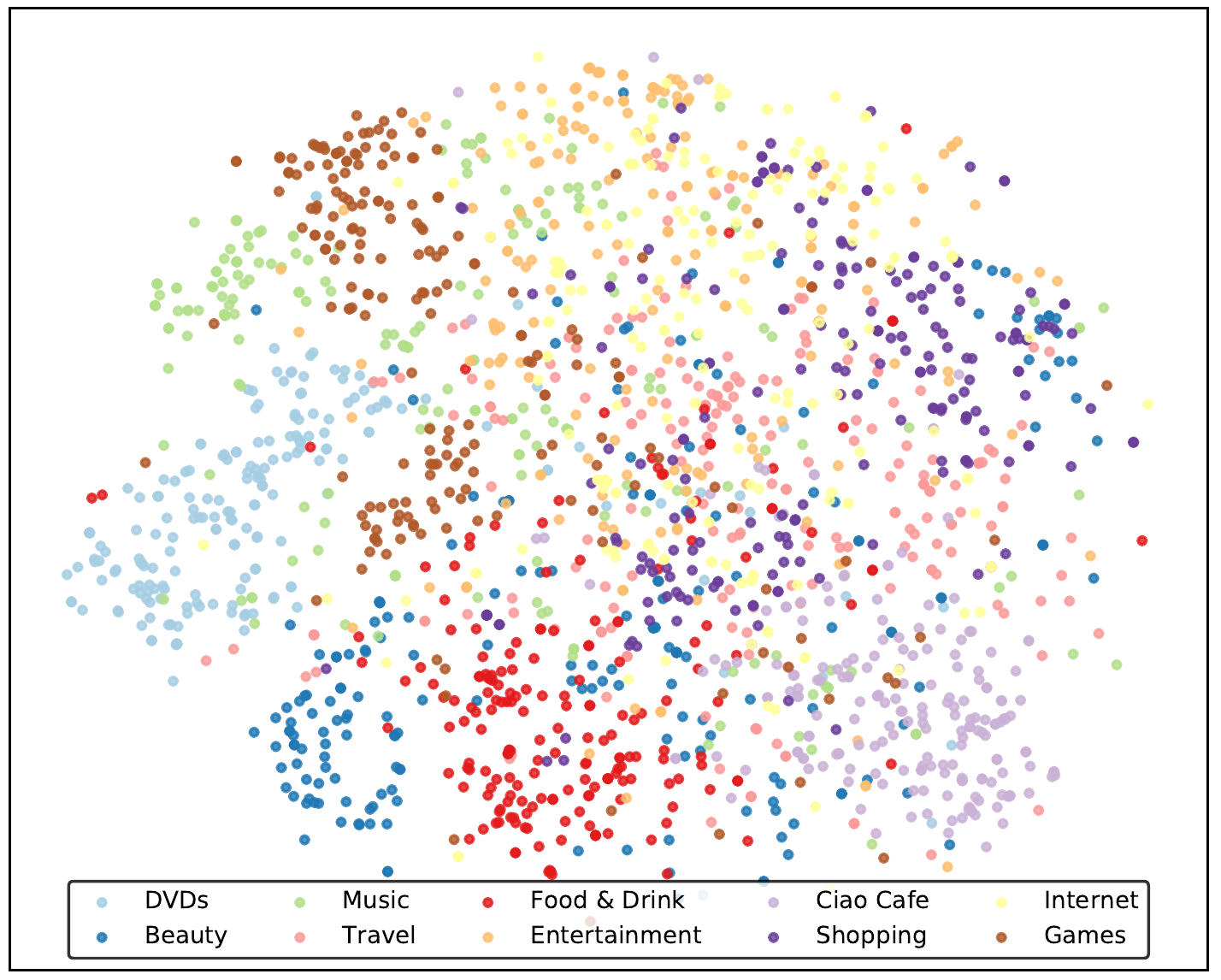}
			\caption{Visualization of $\bm{r}^{\propose}_{ui}$}
		\end{subfigure}
		%		\vspace{-2.5ex}
		\caption{t-SNE visualization of translation vectors of Ciao dataset regarding the item categories.}
		\label{fig:tsne}
	\end{minipage}\hfill
	\begin{minipage}{0.4\textwidth}
		\centering
		%		\vspace{0pt}	\includegraphics[width=\textwidth]{./figure/correlation_ciao.pdf}
		\vspace{-7pt}
		\includegraphics[scale=0.4]{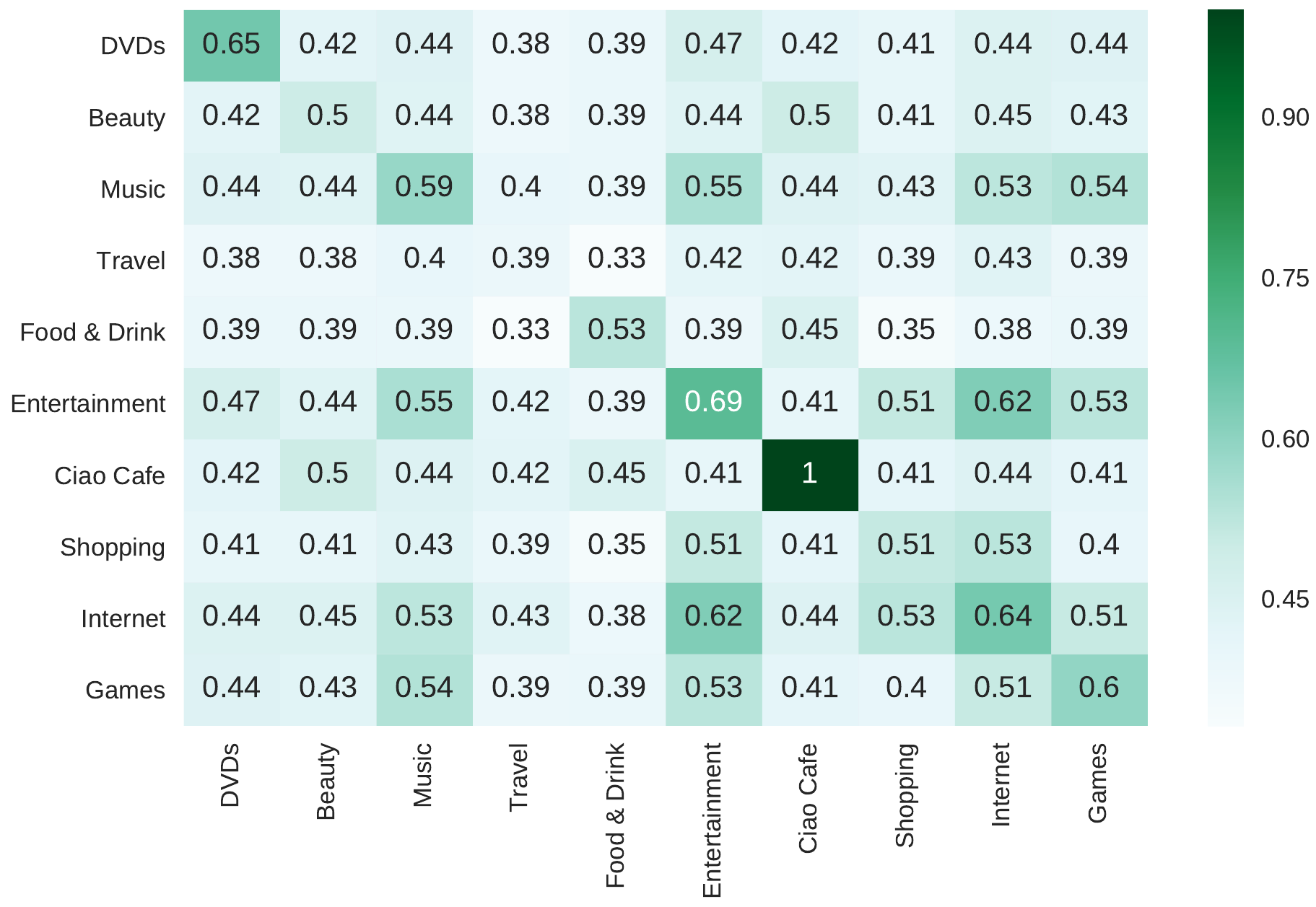}
		%		\vspace{-5.5ex}
		\caption{Heat map of cosine similarity between translation vectors of Ciao dataset. }
		\label{fig:cosine}
	\end{minipage}
	\vspace{-2ex}
\end{figure*}

\smallskip
\noindent\textbf{\textit{Heterogeneity} of user--item interactions. }
In this section, we investigate the translation vectors from a different perspective, i.e., heterogeneity. 
Recall that~\propose~aims to translate each user toward multiple items according to the user's heterogeneous tastes in various item categories, implying that
%tastes for item aspects, such as item genres and categories.
the translation vectors encode information related to item categories.
%are trained so as to 
%encode users' diverse tastes, 
%which makes users with similar tastes have similar translation vectors.
%and that similar translation vectors encode similar information.
%For example, to compute the cosine similarity between translation vectors in category A and B, we sum the cosine similarities of every pair possible pair. i.e., $200\times200$ pairs in our case
%Therefore, we users with similar tastes have similar translation vectors, which implies that users' diverse tastes are encoded in the translation vectors. 
To verify this, we study whether we can conversely infer the categories of items using the translation vectors, \textit{even though the item category information is not utilized during the training of~\propose}. 
To this end, we label each translation vector $\bm{r}_{ui} $ with its corresponding category, and select vectors from the ten most frequently appearing categories
%\footnote{\textbf{Ciao}: [\textquotesingle Ciao Cafe\textquotesingle, \textquotesingle DVDs\textquotesingle, \textquotesingle Internet\textquotesingle, \textquotesingle Food \& Drink\textquotesingle, \textquotesingle Beauty\textquotesingle, \textquotesingle Games\textquotesingle, \textquotesingle Entertainment\textquotesingle, \textquotesingle Travel\textquotesingle, \textquotesingle Music\textquotesingle, \textquotesingle Shopping\textquotesingle], \textbf{Amazon C\&A}: [\textquotesingle Basic Cases\textquotesingle, \textquotesingle Chargers\textquotesingle, \textquotesingle Screen Protectors\textquotesingle, \textquotesingle Headsets\textquotesingle, \textquotesingle Batteries\textquotesingle, \textquotesingle Data Cables\textquotesingle, \textquotesingle Car Accessories\textquotesingle, \textquotesingle Accessory Kits\textquotesingle, \textquotesingle Stylus Pens\textquotesingle, \textquotesingle Unlocked Cell Phones\textquotesingle]} 
to perform classification.
%Given the item category information in Ciao and Amazon C\&A datasets, we 
%and select the translation vectors whose labels belong to the top-10 most frequently appeared categories\footnote{\textbf{Ciao}: ['Ciao Cafe', 'DVDs', 'Internet', 'Food \& Drink', 'Beauty', 'Games', 'Entertainment', 'Travel', 'Music', 'Shopping'], \textbf{Amazon C\&A}: ['Basic Cases', 'Chargers', 'Screen Protectors', 'Headsets', 'Batteries', 'Data Cables', 'Car Accessories', 'Accessory Kits', 'Stylus Pens', 'Unlocked Cell Phones']}. 
%Table~\ref{tab:itemclassification} summarizes the classification results\footref{note1}.
Table~\ref{tab:itemclassification:Trans} shows the item category classification accuracy on the translation vectors\footref{note1}. We observe that~\propose~considerably outperforms CML. This implies that the user--item specific translation vectors indeed encode the \textbf{heterogeneity} of the user--item relationships with regard to users' tastes in various item categories,
%users' diverse tastes for item aspects, 
which provides a justification for the superior performance of~\propose~on Ciao and Amazon C\&A datasets.

We further conduct another experiment on the item embedding vectors $\bm{\beta}_i$ to ascertain that the considerable performance improvement with~\propose~is indeed derived from the translation vectors; not from the high-quality item embedding vectors. This time, we select items whose categories belong to the ten most frequent appearing categories. Table~\ref{tab:itemclassification:Item} shows the classification accuracy on the item embedding vectors $\bm{\beta}_i$ for all $i\in\mathcal{I}$ trained by CML and~\propose. We observe that CML and~\propose~show comparable classification performance, implying that the superior performance of~\propose~is derived not from the high-quality item embedding vectors, but from the translation vectors.

To provide a more intuitive understanding of the numerical results shown in Table~\ref{tab:itemclassification:Trans}, we visualize in Figure~\ref{fig:tsne} the translation vectors learned by CML and~\propose~on Ciao dataset by using t-distributed Stochastic Neighbor Embedding\footnote{We sample 200 samples from each category for visualization.}(t-SNE)~\cite{maaten2008visualizing} with perplexity 30. Each point represents a translation vector, and the color represents the item category. We can clearly see that the translation vectors of~\propose~are generally grouped together according to their corresponding categories, unlike CML; this supports the superior classification results of~\propose~over CML in Table~\ref{tab:itemclassification:Trans}.
%Therefore, we label each translation vector with its corresponding category, and sample 200 vectors from each of the top-10 most frequently appeared categories\footnote{\textbf{Ciao}: ['Ciao Cafe', 'DVDs', 'Internet', 'Food \& Drink', 'Beauty', 'Games', 'Entertainment', 'Travel', 'Music', 'Shopping'], \textbf{Amazon C\&A}: ['Basic Cases', 'Chargers', 'Screen Protectors', 'Headsets', 'Batteries', 'Data Cables', 'Car Accessories', 'Accessory Kits', 'Stylus Pens', 'Unlocked Cell Phones']}.

\begin{table}[t]
%	\small
	\caption{Results of item category classification.}
%	\vspace{-1ex}	
	\begin{subtable}{1\linewidth}
		\centering
		\begin{tabular}{|c||c|c|c|}
			%			\hline
			%			\multicolumn{4}{|c|}{Translation Vector ($\bm{r}_{ui}$)}    \\
			\hline
			Dataset&               Method             & Rand.                   & Random Forest      \\
			\hline
			\multirow{3}{*}{Ciao}& CML               & \tikz[overlay] \draw (0,0.7em)--(0,-0.1em); & 67.86{$\pm0.47$}\% \\
			&  \proposeemb &        10.01\%                  & 67.27{$\pm0.28$}\% \\
			& \propose          &    \tikz[overlay]\draw (0,0.7em)--(0,-0.1em);                      & \textbf{80.97{$\pm0.73$}}\%\\
			\hline
			\multirow{3}{*}{\begin{tabular}[x]{@{}c@{}}Amazon\\C\&A\end{tabular}}& CML               & \tikz[overlay] \draw (0,0.7em)--(0,-0.1em); & 54.26{$\pm0.74$}\% \\
			&  \proposeemb &        10.40\%                  & 54.85{$\pm0.51$}\% \\
			& \propose          &    \tikz[overlay]\draw (0,0.7em)--(0,-0.1em);                      & \textbf{81.24{$\pm0.46$}}\%\\
			\hline
		\end{tabular}
		\medskip
		\caption{Classification on translation vectors ($\bm{r}_{ui}$).}\label{tab:itemclassification:Trans}
	\end{subtable}
	\begin{subtable}{1\linewidth}
		\centering
		\begin{tabular}{|c||c|c|c|}	
			%			\hline
			%			\multicolumn{4}{|c|}{Item Embedding ($\bm{\beta_i}$)}    \\
			\hline
			{Dataset}&        {Method}                    & Rand.                   & Random Forest      \\
			\hline
			\multirow{2}{*}{Ciao}  & CML               & \multirow{2}{*}{10.92\%} & {80.41}{$\pm1.59$}\% \\
			& \propose          &                          & {81.61}{$\pm1.54$}\% \\
			\hline
			\multirow{2}{*}{\begin{tabular}[x]{@{}c@{}}Amazon\\C\&A\end{tabular}}  & CML               & \multirow{2}{*}{9.40\%} & {47.94}{$\pm3.34$}\% \\
			& \propose          &                          & {47.90}{$\pm2.54$}\% \\
			\hline
		\end{tabular}
		\medskip
		\caption{Classification on item embeddings ($\bm{\beta_i}$).}\label{tab:itemclassification:Item}
	\end{subtable}
	
	\label{tab:itemclassification}
	\vspace{-5ex}
\end{table}

Lastly, based on the above observation that similar translation vectors encode similar information with regard to item categories,
%In addition to the above observations, 
we further investigate whether the translation vectors might even reveal the correlations among the item categories.
To this end, 
%assuming that similar translation vectors encode similar information regarding item categories.
we compute the sum of the cosine similarity scores between all pairs of translation vectors from each category. 
We use the same sampled translation vectors that were used for the visualization in Figure~\ref{fig:tsne}.
%\footnote{We use the same sampled translation vectors that were used for the visualization in Figure~\ref{fig:tsne}.}
Figure~\ref{fig:cosine} shows the heat map of cosine similarity between translation vectors learned by~\propose~on Ciao dataset with regard to the item categories. The number in each cell denotes the average cosine similarity score normalized by the overall largest score.
We observe that the similarity score is generally the highest within the same item category (the diagonal line). Moreover, interestingly, semantically related categories also show high correlations. For example, ``Entertainment'' is highly correlated with ``Internet'', ``Games'', ``Music'', and ``Shopping'' all of which are related to entertainment.
%and ``Food\&Drink'' is relatively highly correlated with ``Ciao Cafe'' which 
From the above analyses, we can conclude that~\propose~can even determine the correlations among the item categories by encoding the heterogeneity of user--item relationships into the translation vectors, which again helps explain the superior performance of~\propose. 
This experiment also verifies our assumption illustrated in Figure~\ref{fig:explanation} that the angles between the translation vectors reflect the heterogeneity of the user--item relationships regarding the user's tastes in various item categories.

\vspace{-1ex}
\section{Related Work}
\noindent{\textbf{Recommendations for Implicit Feedback. }}
Although explicit feedback such as rating or review comment is a valuable source of information that reveals user preferences, in most cases it is difficult to obtain a large quantity of such data. Hence, the vast majority of past research has focused on recommendations for implicit feedback~\cite{weston2011wsabie,pan2013gbpr,lee2014local,johnson2014logistic,rendle2009bpr,pan2008one,he2016fast,he2016vbpr}. These methods generally adopt the matrix factorization (MF) technique, which uses the inner product to model the similarity of user--item pairs~\cite{koren2008factorization}. However, the inner product violates the triangle inequality, and thus it may fail to capture fine-grained user preferences~\cite{hsieh2017collaborative}. 

Incorporating the neighborhood information~\cite{desrosiers2011comprehensive} into CF has been shown to be effective for memory--based~\cite{sarwar2001item} and model--based~\cite{ning2011slim,kabbur2013fism,wu2016collaborative,koren2008factorization} CF. 
ItemCF~\cite{sarwar2001item} computes the similarity scores, such as Pearson Correlation and Cosine Similarity, between users based on the items they interacted with in the past.
SLIM~\cite{ning2011slim} and FISM~\cite{kabbur2013fism} improve ItemCF by learning the item-item similarity directly from the data through factorizing the item similarity matrix as a product of two latent factor matrices. CDAE~\cite{wu2016collaborative} is the state-of-the-art neighborhood--based CF method implemented by using denoising auto-encoder. However, previously proposed neighborhood--based CF methods are generally based on MF, and thus suffer from the violation of triangle inequality.

In the following, we briefly introduce metric learning approaches that address this limitation of MF.
%\textcolor{red}{TODO: More related work}

\smallskip
\noindent{\textbf{Metric Learning. }}
Metric learning~\cite{yang2006distance} learns a distance metric that preserves the distance relation among the training data; i.e., it assigns shorter distances to semantically similar pairs. It has been popularized in various domains such as computer vision~\cite{bellet2013survey} and natural language processing~\cite{lebanon2006metric}. Recently, metric learning approaches have been adopted in collaborative filtering to address the limitation of MF.
%~\cite{khoshneshin2010collaborative,chen2012playlist,feng2015personalized,hsieh2017collaborative}. 
Khoshneshin and Street firstly introduced the Euclidean embedding scheme for explicit feedback-based CF, a scheme in which users and items are embedded according to their Euclidean similarity rather than their inner product~\cite{khoshneshin2010collaborative}. For music recommendation, Chen~\etal represent songs as points in Euclidean space, and model the transition probability based on the Euclidean distance between songs~\cite{chen2012playlist}. For point-of-interest (POI) recommendation, Feng~\etal model personalized check-in sequences by projecting each POI into one object in Euclidean space~\cite{feng2015personalized}. 
The above methods can be subsumed by the recently proposed CML~\cite{hsieh2017collaborative}, which is a general recommendation framework for implicit feedback. CML projects users and items into a common Euclidean space in which the similarity of a user--item pair is computed based on the Euclidean distance between the latent vectors of the user and of the item. Given user $u$ and item $i$, the scoring function $s(u,i)$ of CML is computed by:
\begin{math}
\small
\label{eqn:CML_score}
s(u,i) = -\norm{\bm{\alpha}_u - {\bm{\beta}_i}}_2^2.
\end{math}
%where $\bm{\alpha}_u$ and $\bm{\beta}_i$ are the embedding vectors of user $u$ and item $i$, respectively.
In spite of its state-of-the-art performance, CML projects each user to a single point, and thus it does not suffice for modeling the intensity and the heterogeneity of the user--item relationships in implicit feedback.

\smallskip
\noindent{\textbf{Knowledge Graph Embedding. }}
\label{rel:Trans}
Knowledge graph embedding refers to the projection of entities and relations in knowledge graphs into low-dimensional vector spaces. Among various knowledge graph embedding methods, translation--based methods~\cite{bordes2013translating,lin2015learning} have shown state-of-the-art performance and scalability compared with traditional factorization--based embedding methods~\cite{nickel2011three,jenatton2012latent}. Recently, the translation mechanism has been adopted for recommendation algorithms. Zhang~\etal~\cite{zhang2016collaborative} integrate collaborative filtering and a knowledge base
for recommendation by adopting TransR~\cite{lin2015learning} to extract the structural knowledge of items from knowledge graphs. He~\etal~\cite{he2017translation} embed items into a transition space, where each user is modeled by a translation vector to model the transition from the previously consumed item to the next item. However, our work is distinguished from the above methods in that we adopt the translation mechanism to model the latent relationships of implicit user--item interactions rather than to extract some knowledge from knowledge graphs, and we do not model the temporal information. Hence, we do not compare them with~\propose~in our experiments.

\section{Conclusion}
Each implicit user--item interaction encodes a different intensity of user--item relationship, and the relationship is heterogeneous regarding the user's taste in different item categories.
In this paper, we propose a novel metric learning--based recommendation method called~\propose~that captures not only the intensity and the heterogeneity of user--item relationships in implicit feedback, but also the complex nature of CF, all of which have been overlooked by previous metric learning--based recommendation approaches.~\propose~employs the neighborhood information of users and items to construct translation vectors whereby a user is translated toward items according to his relationships with the items.
% while the user is translated farther away from his negative items
Through extensive experiments, we demonstrate that~\propose~considerably outperforms several state-of-the-art methods by generating meaningful translation vectors. 
Regarding possible directions for future studies, refer to Section~\ref{sec:discussion}.

\noindent{\textbf{Acknowledgment:}}
NRF-2016R1E1A1A01942642, NRF-2017M3C4A7063570, IITP-2018-2011-1-00783, IITP-2018-0-00584, and SKT.

%\begingroup
%\setstretch{0.9}
\vspace{-3ex}
\bibliographystyle{IEEEtran}{	
	\bibliography{conference_041818}
}
%\endgroup
\end{document}